\documentclass[%
 aip,
 amsmath,amssymb,
 reprint,%
]{revtex4-1}

\usepackage{graphicx}
\usepackage{dcolumn}
\usepackage{bm}
\usepackage[utf8]{inputenc}
\usepackage[T1]{fontenc}
\usepackage{mathptmx}
\usepackage{color}

\begin{document}

\preprint{AIP/123-QED}


\title[ ]{Raman spectroscopy of group-IV Ge$_{1-x}$Sn$_{x}$ alloys: theory and experiment}

\author{Daniel S.~P.~Tanner}
\thanks{Co-first author}
\affiliation{Laboratoire SPMS, CNRS-CentraleSup\'{e}lec, Universit\'{e} Paris-Saclay, 91190 Gif-sur-Yvette, France}
\affiliation{Physique Th\'{e}orique des Mat\'{e}riaux, CESAM, Universit\'{e} de Li\`{e}ge, B-4000 Li\`{e}ge, Belgium}

\author{Sreyan Raha}
\thanks{Co-first author}
\affiliation{Department of Physics, Bose Institute, 93/1, Acharya Prafulla Chandra Road, Kolkata 700 009, India}

\author{Jessica Doherty}
\affiliation{School of Chemistry, University College Cork, Cork T12 YN60, Ireland}
\affiliation{Tyndall National Institute, University College Cork, Lee Maltings, Dyke Parade, Cork T12 R5CP, Ireland}
\affiliation{AMBER Centre, Environmental Research Institute, University College Cork, T23 XE10, Ireland}

\author{Subhajit Biswas}
\affiliation{School of Chemistry, University College Cork, Cork T12 YN60, Ireland}
\affiliation{Tyndall National Institute, University College Cork, Lee Maltings, Dyke Parade, Cork T12 R5CP, Ireland}
\affiliation{AMBER Centre, Environmental Research Institute, University College Cork, T23 XE10, Ireland}

\author{Justin D.~Holmes}
\affiliation{School of Chemistry, University College Cork, Cork T12 YN60, Ireland}
\affiliation{Tyndall National Institute, University College Cork, Lee Maltings, Dyke Parade, Cork T12 R5CP, Ireland}
\affiliation{AMBER Centre, Environmental Research Institute, University College Cork, T23 XE10, Ireland}

\author{Eoin P.~O'Reilly}
\affiliation{Tyndall National Institute, University College Cork, Lee Maltings, Dyke Parade, Cork T12 R5CP, Ireland}
\affiliation{Department of Physics, University College Cork, Cork T12 YN60, Ireland}

\author{Achintya Singha}
\affiliation{Department of Physics, Bose Institute, 93/1, Acharya Prafulla Chandra Road, Kolkata 700 009, India}

\author{Christopher A.~Broderick}
\email{c.broderick@umail.ucc.ie}
\affiliation{Tyndall National Institute, University College Cork, Lee Maltings, Dyke Parade, Cork T12 R5CP, Ireland}
\affiliation{Department of Physics, University College Cork, Cork T12 YN60, Ireland}

\date{\today}


\begin{abstract}

Ge$_{1-x}$Sn$_{x}$ alloys are a promising candidate material to realise direct-gap group-IV semiconductors for applications in Si-compatible electronic and photonic devices. Here, we present a combined theoretical and experimental analysis of Raman spectroscopy in Ge$_{1-x}$Sn$_{x}$ alloys. We describe liquid-vapour-solid growth and structural characterisation of Ge$_{1-x}$Sn$_{x}$ ($x \leq 8$\%) nanowires displaying high crystalline quality, and investigate the structural and vibrational properties of the nanowires using Raman spectroscopy. Our theoretical analysis is based on a fully analytic anharmonic valence force field (VFF) potential, which describes exactly -- i.e.~without recourse to numerical fitting -- the second-order elastic constants, third-order bulk modulus, selected second- and third-order inner elastic constants and, as a consequence, the zone-centre transverse optical phonon mode frequency and its hydrostatic and axial strain dependence. We compute bulk elastic properties via density functional theory to parametrise the VFF potential for Ge$_{1-x}$Sn$_{x}$ alloys, and apply the VFF potential to explicitly compute the Raman spectra of realistic, disordered Ge$_{1-x}$Sn$_{x}$ alloy supercells. Our atomistic theoretical calculations quantitatively capture: (i) the evolution of the measured Raman spectra with Sn composition $x$, (ii) demonstrate explicitly that the presence of short-range alloy disorder can significantly impact the shift coefficients $a$ and $b$ that respectively describe the dependence of the Raman shift on Sn composition and pseudomorphic strain, (iii) elucidate the origin of the so-called ``disorder activated'' mode identified in previous experimental investigations, and (iv) allow for detailed atomic-scale interpretation of measured Raman spectra. Overall, our analysis provides insight relevant to the characterisation of this emerging material system.

\end{abstract}

\maketitle


\section{Introduction}
\label{sec:introduction}


The emergence of a direct band gap in Ge$_{1-x}$Sn$_{x}$ alloys for low Sn compositions $x \approx 6$ -- 9\% has stimulated significant research interest in this isovalent group-IV material system as a potential route to obtain direct-gap semiconductor materials compatible with conventional complementary metal-oxide-semiconductor (CMOS) fabrication and processing infrastructure. \cite{Kouvetakis_ARMR_2006,Soref_NP_2010,Kasper_PR_2013,Soref_PTRSA_2014,Geiger_FM_2015,Reboud_PCGC_2017,Doherty_CM_2020,Moutanabbir_APL_2021} In recent years, this interest has motivated research aimed at exploiting the novel electronic, optical and transport properties of (Si)Ge$_{1-x}$Sn$_{x}$ alloys and nanostructures for a range of applications, including in active and passive components for integrated Si photonics, \cite{Wirths_NP_2015,Zhou_Optica_2020,Su_OE_2011,Peng_APL_2014,Tran_ACSP_2019,Chen_JAP_2021} as well as in electronics \cite{Yang_IEEETED_2013,Wirths_APL_2013,Wirths_ACSAMI_2015,Schulte-Braucks_SSE_2017,Schutte-Braucks_IEEETED_2017} and photovoltaics. \cite{Fang_JACS_2008,Beeler_ECSJSSST_2013,Ventura_PIPRA_2015,Roucka_IEEEJPV_2016,Pearce_SiGeSn_2021} Key to the realisation of these proposed semiconductor devices is the ability to reproducibly grow high-quality (Si)Ge$_{1-x}$Sn$_{x}$ alloys and nanostructures. This mandates the availability of reliable structural and chemical characterisation techniques, supported by quantitative atomic-scale understanding of material properties and their impact on measurable physical properties. Due to its contactless and non-destructive nature, Raman spectroscopy has been widely employed to characterise bulk-like Ge$_{1-x}$Sn$_{x}$ alloys and Ge$_{1-x}$Sn$_{x}$-based nanostructures, \cite{Rojas-Lopez_JAP_1998,Li_APL_2004,Costa_SSC_2007,Lin_APL_2011,Su_SSC_2011,Bagchi_PRB_2011,Oehme_JCG_2013,Cheng_ECSJSSST_2013,Fournier-Lupien_APL_2013,Lieten_ECSJSSST_2014,Dash_CEC_2014,Chang_TSF_2015,Lei_APL_2016,Takeuchi_JJAP_2016,Gassenq_APL_2017,Perova_JRS_2017,Xu_APL_2019,Bouthillier_SST_2020,Liu_JRS_2020,Wang_IEEETED_2020,Schlipf_JRS_2021} where it can provide valuable insight into the crystal structure, compositional homogeneity, and vibrational (phononic) properties. To date, almost all consideration of Raman scattering in Ge$_{1-x}$Sn$_{x}$ alloys has been from an experimental perspective, and has centred on empirical parameter extraction from measured Raman spectra. Characterisation of Ge$_{1-x}$Sn$_{x}$ alloys via Raman spectroscopy is in principle capable of providing detailed insight into the structural and vibrational properties. However, in practice the lack of a predictive atomic-scale approach to simulate and interrogate Ge$_{1-x}$Sn$_{x}$ Raman spectra restricts the ability to draw conclusions regarding these properties to their high-level dependence on alloy composition and strain, rather than on the details of the alloy microstructure. In this work we aim to close this gap, by establishing an efficient and accurate atomistic theoretical framework to compute Raman spectra in Ge$_{1-x}$Sn$_{x}$ alloys, thereby providing an enhanced level of insight to support this commonly-employed characterisation technique.


Here, we present a combined theoretical and experimental analysis of Raman activity in Ge$_{1-x}$Sn$_{x}$ alloy nanowires (NWs). We perform Raman spectroscopy measurements on high-quality Ge$_{1-x}$Sn$_{x}$ NWs ($x \leq 8$\%) fabricated using a non-equilibrium vapour-liquid-solid (VLS) approach. The measurements demonstrate the well known redshift and broadening of the Raman feature associated with Ge-Ge optical phonon modes with increasing Sn composition $x$. To interpret these measurements we develop a theoretical model based on an anharmonic valence force field (VFF) potential, which is used to perform lattice relaxation and compute the phonon modes in disordered Ge$_{1-x}$Sn$_{x}$ alloy supercells, with these phonon modes then used explicitly to compute alloy Raman spectra. The anharmonic VFF potential is based closely on the harmonic VFF potential we have recently developed to enable efficient and accurate lattice relaxation of group-IV alloys and nanostructures. \cite{Tanner_VFF_2021} The harmonic VFF potential of Ref.~\onlinecite{Tanner_VFF_2021} was parametrised analytically by expressing the force constants of the potential in terms of the relaxed second-order elastic constants $C_{ij}$ and the internal strain (Kleinman) parameter $\zeta$, so that these linear elastic properties are exactly reproduced by the potential without recourse to numerical fitting. Here, to better capture the harmonic lattice dynamical properties, we effect an adjusted analytic parametrisation to use the second-order inner elastic constant $B_{xx}$ in place of $\zeta$, so that the harmonic terms of the potential exactly reproduce $C_{ij}$ and the zone-centre transverse optical (TO) phonon frequency $\omega_{\scalebox{0.7}{TO}}$. The anharmonic terms of the potential are then chosen in order to exactly reproduce the third-order (anharmonic) bulk modulus and selected third-order inner elastic constants, retaining an analytical parametrisation that exactly describes the dependence of $\omega_{\scalebox{0.7}{TO}}$ on applied hydrostatic and axial strains. We compute the complete set of structural and elastic properties required to analytically parametrise the anharmonic VFF potential from first principles --  using density functional theory (DFT) -- for the constituent materials Ge, $\alpha$-Sn and the zinc blende IV-IV compound GeSn (zb-GeSn), to determine the full set of VFF parameters required to perform atomistic alloy supercell calculations.

Our high-throughput alloy supercell calculations then describe quantitatively the experimentally observed redshift and asymmetric frequency broadening of the Raman feature associated with Ge-Ge optical phonon modes. We elucidate the atomic-scale origin of these behaviours in terms of: (i) phonon mode softening due to global lattice expansion driven by Sn incorporation, which acts to lower the optical vibrational frequencies of Ge-Ge bonds, and (ii) the role played by the alloy microstructure, where local lattice relaxation driven by short-range chemical inhomogeneity produces a distribution of closely-spaced Ge-Ge bond optical vibrational frequencies. In the existing literature, the dependence of the shift $\Delta \omega$ of the Raman activity-weighted average Ge-Ge mode frequency on Sn composition $x$ and in-plane strain $\epsilon_{\parallel}$ are respectively described using the shift coefficients $a$ and $b$ (cf.~Eq.~\eqref{eq:mode_frequency_shift}). These parameters have to date been estimated for Ge$_{1-x}$Sn$_{x}$ alloys either by fitting to measured $\Delta \omega$ data, or by simple linear interpolation between the known properties of Ge and $\alpha$-Sn, introducing systematic uncertainties and leading to a large spread of reported values. \cite{Costa_SSC_2007} Using atomistic calculations on free-standing (relaxed) and pseudomorphically strained alloy supercells we extract $a$ and $b$ explicitly, demonstrating good quantitative agreement with our experimental measurements and with literature data. Additional comparative calculations for ordered and disordered alloy supercells highlight that the presence of short-range alloy disorder acts to reduce the magnitude of $a$ and $b$ compared to the values computed for idealised ordered alloys. Finally, for one of the NWs investigated ($x = 6$\%), our measurements reveal a prominent ``shoulder'' on the high-frequency side of the Ge-Ge Raman feature. Theoretical analysis of the structural and vibrational properties of disordered alloy supercells identifies that this shoulder can be associated with the presence of excess Sn-Sn nearest-neighbour bonds compared to the number expected in a randomly disordered Ge$_{1-x}$Sn$_{x}$ alloy, and can then act as a spectroscopic fingerprint of Sn clustering in real alloy samples.


The remainder of this paper is organised as follows. We describe our experimental methodology in Sec.~\ref{sec:experiment}, including details of the sample growth and characterisation in Sec.~\ref{sec:experiment_samples}, and Raman spectroscopic measurements in Sec.~\ref{sec:experiment_raman}. Our theoretical methodology is described in Sec.~\ref{sec:theory}, beginning in Sec.~\ref{sec:theory_dft} with the details of the DFT calculations, before describing in Sec.~\ref{sec:theory_vff} the anharmonic VFF potential and its analytical parametrisation. We then outline our atomistic calculation of Raman spectra in Sec.~\ref{sec:theory_raman}. The results of our combined theoretical and experimental investigation are presented in Sec.~\ref{sec:results}, beginning in Sec.~\ref{sec:results_composition} with the Sn composition dependence of the Ge-Ge mode Raman shift. Following this, in Sec.~\ref{sec:results_disorder} we describe the impact of alloy disorder on the Ge-Ge mode Raman shift, before considering in Sec.~\ref{sec:results_strain} the impact of pseudomorphic strain. Then, in Sec.~\ref{sec:results_microstructure} we consider the impact of the alloy microstructure on the Raman spectrum. Finally, in Sec.~\ref{sec:conclusions} we summarise and conclude.


\section{Experimental methods}
\label{sec:experiment}


\subsection{Sample growth and characterisation}
\label{sec:experiment_samples}


\begin{figure*}[t!]
	\includegraphics[width=0.95\textwidth]{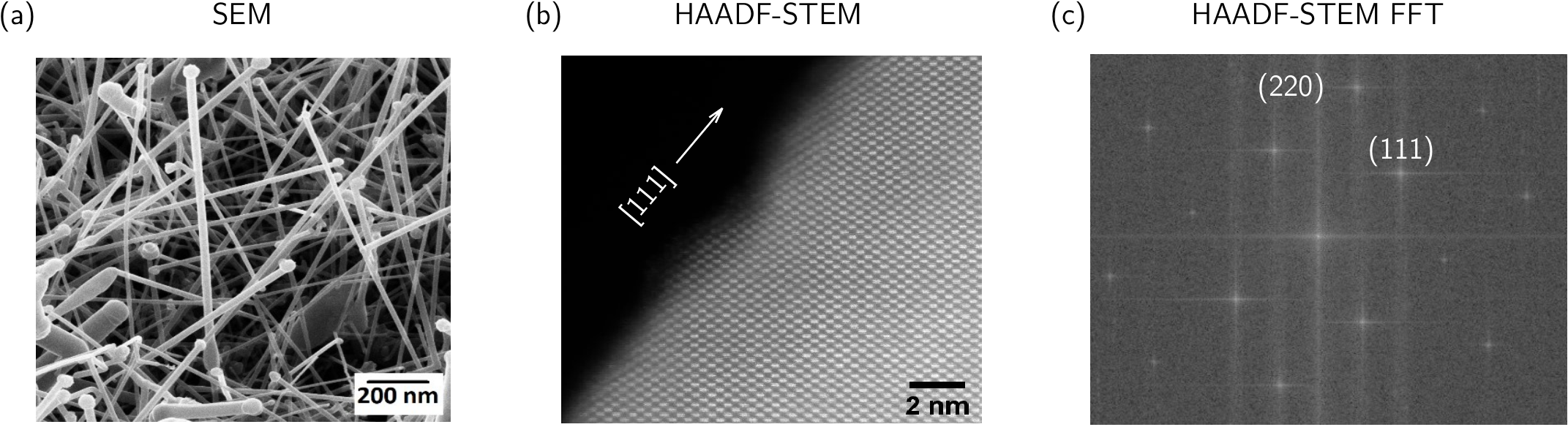}
	\caption{Structural characterisation of the LICVD-grown Ge$_{1-x}$Sn$_{x}$ NW sample having Sn composition $x = 8$\%, including (a) scanning electron microscopy (SEM), (b) high-angle annular dark-field scanning transmission electron microscopy (HAADF-STEM), and (c) the fast Fourier transform (FFT) of (b). The STEM measurements of (b) and (c) confirm high-quality, single crystal (diamond-structured) NW growth.}
	\label{fig:structural_characterisation}
\end{figure*}

Ge$_{1-x}$Sn$_{x}$ NWs were grown via a bottom-up method using a liquid-injection chemical vapour deposition (LICVD) technique. \cite{Biswas_NC_2016,Doherty_JMCC_2018,Doherty_CM_2019} Following the synthesis method described in Ref.~\onlinecite{Biswas_NC_2016}, Ge$_{1-x}$Sn$_{x}$ NWs were grown on Si substrates using diphenylgermane (C$_{12}$H$_{12}$Ge; DPG) and allyltributylstanane (C$_{15}$H$_{32}$Sn; ATBS) as the Ge and Sn precursors, respectively. Growth was performed at a temperature of 440 $^{\circ}$C using Au and Au$_{0.9}$Ag$_{0.1}$ nanoparticles as catalysts. Three Ge$_{1-x}$Sn$_{x}$ NW samples, with Sn compositions $x = 1$, 6 and 8\% were grown by varying the concentration of ATBS in the precursor solution. The structural properties of the Ge$_{1-x}$Sn$_{x}$ NWs were characterised via a combination of scanning electron microscopy (SEM) and high-resolution scanning transmission electron microscopy (STEM), the latter performed in high-angle annular dark-field mode (HAADF-STEM). The SEM measurements were performed using a Zeiss ORION NanoFab microscope, while HR-STEM imaging was performed using a Nion UltraSTEM100 microscope operated at an accelerating voltage of 100 kV. To confirm the NW Sn compositions, energy-dispersive x-ray (EDX) mapping measurements were performed in HAADF mode, using a FEI Helios NanoLab 600i system operated at an accelerating voltage of 30 kV and a probe current of 0.69 nA, with detection via an attached Oxford Instruments X-Max 80 spectrometer. Results of the EDX measurements, as well as further details of the growth and characterisation of the NWs investigated in this work, can be found in Refs.~\onlinecite{Biswas_NC_2016}, \onlinecite{Doherty_JMCC_2018} and \onlinecite{Doherty_CM_2019}.

Figure~\ref{fig:structural_characterisation}(a) shows a representative SEM image of the Ge$_{0.92}$Sn$_{0.08}$ ($x = 8$\%) NW sample. The SEM image indicates uniform NW morphology, demonstrating negligible presence of metallic Sn clusters in the sample. \cite{Biswas_NC_2016} This high quality crystalline growth is representative of all bottom-up LICVD-grown Ge$_{1-x}$Sn$_{x}$ NWs investigated in this work. The NW diameters are in the 40 -- 60 nm range, with the NW lengths ranging from 1 -- 2 $\mu$m. The average Sn compositions in the NWs, determined via EDX analysis, were $x = 1$, 6 and 8\% for the samples investigated. EDX mapping confirms a homogeneous distribution of Sn in the NW bodies, with a Sn-rich nanoparticle at the NW tip. \cite{Doherty_JMCC_2018} Figure~\ref{fig:structural_characterisation}(b) shows a representative STEM image for a single Ge$_{0.92}$Sn$_{0.08}$ NW, confirming the high crystalline quality. Similar structural properties were observed via STEM imaging of the $x = 1$ and 6\% NW samples (not shown). For all three NW samples, no extended crystalline defects -- e.g.~stacking faults or twinning -- were observed. From the STEM imaging of the $x = 8$\% NWs in Fig.~\ref{fig:structural_characterisation}(b), an interplanar spacing $d = 0.340$ nm was inferred. This interplanar spacing is marginally larger than the value $d = 0.326$ nm in bulk Ge, corresponding to the separation between atomic layers in the [111] orientation of the cubic (diamond) structure, reflecting the expected lattice relaxation to a larger alloy lattice parameter due to Sn incorporation. Indeed, the recorded fast Fourier transform (FFT) of the STEM image in Fig.~\ref{fig:structural_characterisation}(c) for the Ge$_{0.92}$Sn$_{0.08}$ NW corresponds closely to that of cubic Ge, with the spot pattern confirming the single-crystal structure of the LICVD-grown Ge$_{1-x}$Sn$_{x}$ NWs. The recorded FFTs displayed pseudo-hexagonal symmetry consistent with the expected common [111] NW growth direction.


\subsection{Experimental measurements: Raman spectroscopy}
\label{sec:experiment_raman}

Raman scattering measurements were performed using a LabRAM HR800 high-resolution confocal micro-Raman spectrometer, equipped with a 1800 line mm$^{-1}$ grating and a Peltier-cooled charge-coupled device detector. An air-cooled argon-ion laser of wavelength 488 nm was employed as the excitation light source, and a 100$\times$ objective lens having a numerical aperture of 0.9 was used to focus the incident laser beam onto the NW samples, and to collect light scattered from the samples. The spot diameter of the incident laser beam at the sample surface was $\approx 0.7$ $\mu$m. The incident beam direction was set perpendicular to the [111] NW growth direction, with the Raman scattered (normal Stokes) light collected along the same direction. All Raman spectroscopic measurements were performed at fixed temperature $T = 83$ K and laser power $P = 0.2$ mW, with temperature control provided via a Linkam THMS600 microscope stage. We direct the reader to Ref.~\onlinecite{Raha_GeSn_Raman_2021} for details of additional temperature- and incident power-dependent Raman spectroscopic measurements for the same set of Ge$_{1-x}$Sn$_{x}$ NW samples investigated in this work.


\section{Theoretical methods}
\label{sec:theory}

Our theoretical analysis of the vibrational properties of Ge$_{1-x}$Sn$_{x}$ alloys is based on an atomistic supercell approach, where large (512-atom) supercells are employed to accommodate the variety of distinct local microstructures that can occur in realistic disordered alloys. \cite{Rucker_PRB_1996} To allow high-throughput atomic relaxation and computation of the vibrational modes for large supercells, it is necessary to employ an accurate empirical interatomic potential to overcome the system size limitations imposed by the computational expense associated with first principles calculations. We describe the establishment of such a potential for Ge$_{1-x}$Sn$_{x}$ in this section, beginning in Sec.~\ref{sec:theory_dft} with DFT calculations of the elastic properties of Ge, $\alpha$-Sn and zb-GeSn, which are then used in Sec.~\ref{sec:theory_vff} to parametrise a fully analytic anharmonic VFF potential that we apply to relax and compute the vibrational modes of Ge$_{1-x}$Sn$_{x}$ alloy supercells. In Sec.~\ref{sec:theory_raman} we describe how the computed alloy supercell vibrational modes are used to compute Raman spectra suitable for direct comparison to experimental measurements.


\subsection{Density functional theory calculations}
\label{sec:theory_dft}

Written in terms of the second-order (harmonic) unrelaxed and inner (``bare'') elastic tensors $C_{ij}^{(0)}$ and $B_{ij}$, and the third-order (anharmonic) unrelaxed and inner elastic tensors $C_{ijk}^{(0)}$ and $B_{ijk}$, the elastic free energy density of a cubic crystal is given, in Voigt notation, by \cite{Vanderbilt_PRB_1989}

\begin{widetext}
    \begin{eqnarray}
        U &=& \frac{1}{2} \, C_{11}^{(0)} \, \bigg( \eta_{1}^{2} + \eta_{2}^{2} + \eta_{3}^{2} \bigg) + C_{12}^{(0)} \, \bigg( \eta_{1} \eta_{2} + \eta_{2} \eta_{3} + \eta_{2} \eta_{3} \bigg) + \frac{1}{2} \, C_{44}^{(0)} \bigg( \eta_{4}^{2} + \eta_{5}^{2} + \eta_{6}^{2} \bigg) + B_{4x} \bigg( t_{x} \, \eta_{4} + t_{y} \, \eta_{5} + t_{z} \, \eta_{6} \bigg) \nonumber \\
        &+& \frac{1}{2} \, B_{xx} \, \bigg( t_{x}^{2} + t_{y}^{2} + t_{z}^{2} \bigg) + \frac{1}{6} \, C_{111}^{(0)} \, \bigg( \eta_{1}^{3} + \eta_{2}^{3} + \eta_{3}^{3} \bigg) + \frac{1}{2} \, C_{112}^{(0)} \, \bigg( \eta_{1}^{2} \left( \eta_{2} + \eta_{3} \right) + \eta_{2}^{2} \left( \eta_{1} + \eta_{3} \right) + \eta_{3}^{2} \left( \eta_{1} + \eta_{2} \right) \bigg) \nonumber \\
        &+& C_{123}^{(0)} \, \eta_{1} \, \eta_{2} \, \eta_{3} + \frac{1}{2} \, C_{144}^{(0)} \, \bigg( \eta_{1} \, \eta_{4}^{2} + \eta_{2} \, \eta_{5}^{2} + \eta_{3} \, \eta_{6}^{2} \bigg) + \frac{1}{2} \, C_{155}^{(0)} \, \bigg( \eta_{1} \left( \eta_{5}^{2} + \eta_{6}^{2} \right) + \eta_{2} \left( \eta_{4}^{2} + \eta_{6}^{2} \right) + \eta_{3} \left( \eta_{4}^{2} + \eta_{5}^{2} \right) \bigg) \nonumber \\
        &+& C_{456}^{(0)} \, \eta_{4} \, \eta_{5} \, \eta_{6} + B_{14x} \, \bigg( \eta_{1} \, \eta_{4} \, t_{x} + \eta_{2} \, \eta_{5} \, t_{y} + \eta_{3} \, \eta_{6} \, t_{z} \bigg) + B_{45z} \, \bigg( \eta_{4} \, \eta_{5} \, t_{z} + \eta_{4} \, \eta_{6} \, t_{y} + \eta_{5} \, \eta_{6} \, t_{x} \bigg) + B_{xyz} \, t_{x} \, t_{y} \, t_{z} \nonumber \\
        &+& B_{15y} \, \bigg( ( \eta_{1} + \eta_{2} ) \, \eta_{6} \, t_{z} + ( \eta_{1} + \eta_{3} ) \, \eta_{5} \, t_{y} + ( \eta_{2} + \eta_{3} ) \, \eta_{4} \, t_{x} \bigg) + \frac{1}{2} \, B_{1xx} \, \bigg( \eta_{1} \, t_{x}^{2} + \eta_{2} \, t_{y}^{2} + \eta_{3} \, t_{z}^{2} \bigg) \nonumber \\
        &+& B_{4yz} \, \bigg( \eta_{6} \, t_{y} \, t_{z} + \eta_{5} \, t_{x} \, t_{z} + \eta_{6} \, t_{x} \, t_{y} \bigg) + \frac{1}{2} \, B_{1yy} \, \bigg( ( \eta_{1} + \eta_{3} ) \, t_{y}^{2} + ( \eta_{1} + \eta_{2} ) \, t_{z}^{2} + ( \eta_{2} + \eta_{3} ) \, t_{x}^{2} \bigg) \, ,
        \label{eq:elastic_energy_unrelaxed}
    \end{eqnarray}
\end{widetext}

\noindent
where $\eta_{i}$ are the components of the finite (Lagrangian) strain tensor and $t_{i}$ are the components of the internal strain vector, which respectively describe macroscopic (acoustic) lattice deformation and internal (optical) lattice displacement.


\begin{table*}[t!]
	\caption{\label{tab:elastic_parameters} LDA-DFT calculated equilibrium lattice parameter $a_{0}$, and harmonic and anharmonic elastic properties of Ge, $\alpha$-Sn and zb-GeSn. Harmonic properties include the relaxed elastic constants $C_{11}$, $C_{12}$ and $C_{44}$, bulk modulus $B = \frac{1}{3} ( C_{11} + 2 C_{12} )$, inner elastic constant $B_{xx}$, and zone-centre optical phonon frequency $\omega_{\scalebox{0.7}{\text{TO}}}$ (calculated via Eq.~\eqref{eq:optical_phonon_frequency}). Anharmonic properties include the anharmonic bulk modulus $B' = \frac{2}{9} ( 3 C_{111} + 6 C_{112} + C_{123} )$, inner elastic constants $B_{1xx}$ and $B_{1yy}$, and zone-centre optical mode Gr\"{u}neisen parameter $\gamma_{\scalebox{0.7}{\text{TO}}}$ (calculated via Eq.~\eqref{eq:mode_gruneisen_parameter}).}
    	\begin{ruledtabular}
    		\begin{tabular}{cccccccccccc}
    			         & $a_{0}$ & $C_{11}$ & $C_{12}$ & $C_{44}$ & $B$     & $B_{xx}$         & $\omega_{\scalebox{0.7}{\text{TO}}}$ & $B^{\prime}$ & $B_{1xx}$        & $B_{1yy}$        & $\gamma_{\scalebox{0.7}{\text{TO}}}$ \\
    			         & (\AA)   & (GPa)    & (GPa)    & (GPa)    & (GPa)   & (GPa \AA$^{-2}$) & (cm$^{-1}$)                          & (GPa)        & (GPa \AA$^{-2}$) & (GPa \AA$^{-2}$) &          \\
    			\hline
    			Ge       & 5.647   & 120.23   & 48.03    & 60.19    & 72.10   & 40.41            & 291.6                                & $-$279.61    & $-$87.60         & $-$98.28         & 1.172    \\
       $\alpha$-Sn       & 6.480   &  66.72   & 34.23    & 28.39    & 45.06   & 18.48            & 189.6                                & $-$175.59    & $-$38.53         & $-$51.61         & 1.278    \\
    		 zb-GeSn     & 6.072   &  87.23   & 41.09    & 40.29    & 56.47   & 25.98            & 234.1                                & $-$219.58    & $-$57.03         & $-$70.55         & 1.271    \\
    		\end{tabular}
    	\end{ruledtabular}
\end{table*}

The internal strain \textbf{t} describes rigid displacement between the two sublattices of a diamond- or zinc blende-structured crystal, so that $B_{4x}$ and $B_{xx}$ respectively describe coupling between acoustic deformations and optical displacements, and between purely optical displacements (internal strains). To second order, one obtains the conventional relaxed elastic tensor $C_{ij}$ by minimising the harmonic terms in Eq.~\eqref{eq:elastic_energy_unrelaxed} -- i.e.~terms to second order in products of $\eta_{i}$ and/or $t_{i}$ -- with respect to the internal strain, yielding $\textbf{t} = - \frac{ a_{0} \zeta }{4} ( \eta_{4}, \eta_{5}, \eta_{6} )$, where $a_{0}$ is the relaxed lattice parameter and the Kleinman parameter $\zeta$ is defined as

\begin{equation}
    \zeta = \frac{ 4 \, B_{4x} }{ a_{0} \, B_{xx} } \, .
    \label{eq:kleinman_parameter}
\end{equation}

Substituting this expression for \textbf{t} into the harmonic terms of Eq.~\eqref{eq:elastic_energy_unrelaxed} then allows to compute the conventional relaxed elastic constants $C_{11} = C_{11}^{(0)}$, $C_{12} = C_{12}^{(0)}$, and $C_{44} = C_{44}^{(0)} - \frac{ B_{4x}^{2} }{ B_{xx} }$. \cite{Vanderbilt_PRB_1989,Tanner_VFF_2021} For brevity, we do not recapitulate the relationship between the relaxed third-order elastic constants $C_{ijk}$ and their unrelaxed counterparts $C_{ijk}^{(0)}$. Details of these relationships can be found in Ref.~\onlinecite{Vanderbilt_PRB_1989}.

The $B_{xx}$ term in Eq.~\eqref{eq:elastic_energy_unrelaxed} depends only on internal strain, such that $B_{xx}$ determines the zone-centre transverse optical (TO) phonon frequency $\omega_{\scalebox{0.7}{\textrm{TO}}}$ via

\begin{equation}
    \omega_{\scalebox{0.7}{\textrm{TO}}} = \sqrt{ \frac{ \Omega_{0} \, B_{xx} }{ \mu } } \, ,
    \label{eq:optical_phonon_frequency}
\end{equation}

\noindent
where $\Omega_{0} = \frac{ a_{0}^{3} }{ 4 }$ and $\mu$ are respectively the equilibrium volume and reduced ionic mass of a two-atom primitive unit cell. We note that in non-polar Ge and $\alpha$-Sn the zone-centre TO and longitudinal optical (LO) phonon modes are degenerate, so that $\omega_{\scalebox{0.7}{\textrm{LO}}}$ ($= \omega_{\scalebox{0.7}{\textrm{TO}}}$) is also given by Eq.~\eqref{eq:optical_phonon_frequency}.

Performing similar analysis for the anharmonic terms in Eq.~\eqref{eq:elastic_energy_unrelaxed} allows to obtain the relaxed third-order elastic constants $C_{ijk}$ in terms of the unrelaxed elastic constants $C_{ijk}^{(0)}$ and inner elastic constants $B_{ijk}$. \cite{Vanderbilt_PRB_1989} However, for our purposes our interest in the anharmonic contributions to the elastic free energy density is limited to the extent to which these contributions determine the strain dependence of the zone-centre TO phonon frequency $\omega_{\scalebox{0.7}{\textrm{TO}}}$. As such, it suffices to note that it is possible to compute the mode Gr\"{u}neisen parameter $\gamma_{\scalebox{0.6}{\textrm{TO}}}$ associated with $\omega_{\scalebox{0.7}{\textrm{TO}}}$ by extending the derivation of Eq.~\eqref{eq:optical_phonon_frequency} -- which considers Eq.~\eqref{eq:elastic_energy_unrelaxed} in the presence of an applied purely optical displacement -- to include anharmonic terms, yielding \cite{Cerdeira_PRB_1972,Vanderbilt_PRB_1989}

\begin{equation}
    \gamma_{\scalebox{0.6}{\textrm{TO}}} = - \frac{ \partial ( \log \omega_{\scalebox{0.7}{\textrm{TO}}} ) }{ \partial ( \log \Omega ) } \bigg|_{\Omega = \Omega_{0}} = - \frac{ \left( B_{1xx} + 2 B_{1yy} \right) }{ 6 B_{xx} } \, .
    \label{eq:mode_gruneisen_parameter}
\end{equation}

Similarly, the phonon deformation potential $a_{s,\scalebox{0.6}{\text{TO}}}$ describing the impact of axial strain on $\omega_{\scalebox{0.7}{\textrm{TO}}}$ is given by \cite{Cerdeira_PRB_1972,Vanderbilt_PRB_1989}

\begin{equation}
    a_{s,\scalebox{0.6}{\text{TO}}} = \frac{ B_{1xx} - B_{1yy} }{ B_{xx} } \, ,
    \label{eq:phonon_deformation_potential_axial}
\end{equation}

\noindent
so that the strain dependence of $\omega_{\scalebox{0.7}{\textrm{TO}}}$ is fully determined by the second-order inner elastic constant $B_{xx}$ (which determines $\omega_{\scalebox{0.7}{\textrm{TO}}}$ in the unstrained case), and the third-order inner elastic constants $B_{1xx}$ and $B_{1yy}$ (which, given $\omega_{\scalebox{0.7}{\textrm{TO}}}$ via $B_{xx}$, determine $\gamma_{\scalebox{0.6}{\textrm{TO}}}$ and $a_{s,\scalebox{0.6}{\text{TO}}}$).

We note from Eq.~\eqref{eq:elastic_energy_unrelaxed} that $B_{1xx}$ and $B_{1yy}$ describe the coupling of internal (optical) displacements to macroscopic (acoustic) lattice deformations, \cite{Vanderbilt_PRB_1989} thereby explaining why $B_{xx}$, $B_{1xx}$ and $B_{1yy}$ are the only elastic constants that appear in Eqs.~\eqref{eq:mode_gruneisen_parameter} and~Eq.~\eqref{eq:phonon_deformation_potential_axial}: these second- and third-order inner elastic constants fully encapsulate the dependence of $\omega_{\scalebox{0.7}{\textrm{TO}}}$ on any combination of axial strains, including hydrostatic strain.

In order to parametrise a VFF potential that is capable of accurately relaxing alloy atomic positions, while simultaneously accurately describing strain-dependent optical phonon frequencies, we require accurate descriptions of: (i) the relaxed second-order elastic constants $C_{11}$, $C_{12}$ and $C_{44}$, and hence the harmonic bulk modulus $B = \frac{1}{3} ( C_{11} + 2 \, C_{12} )$, (ii) the harmonic inner elastic constant $B_{xx}$, (iii) the relaxed third-order elastic constants $C_{111}$, $C_{112}$ and $C_{123}$, and hence the anharmonic bulk modulus $B' = \frac{2}{9} ( 3 C_{111} + 6 C_{112} + C_{123} )$, and (iv) the anharmonic inner elastic constants $B_{1xx}$ and $B_{1yy}$. We note that the anharmonic bulk modulus $B'$ is closely related to, but not to be confused with, the pressure derivative of the harmonic bulk modulus $B$.

Determination of inner elastic constants has not generally been the primary focus of experimental investigations of lattice elastic and phononic properties, making it challenging to assemble from the literature a complete and consistent set of second- and third-order inner elastic constants suitable for interatomic potential parametrisation. As such, we compute these parameters explicitly for Ge, $\alpha$-Sn and zb-GeSn using DFT in the local density approximation (LDA). We have recently calculated the equilibrium lattice parameter $a_{0}$, relaxed second-order elastic constants $C_{11}$, $C_{12}$ and $C_{44}$, and second-order inner elastic constant $B_{xx}$ for these materials in Ref.~\onlinecite{Tanner_VFF_2021}. Here, we extend these calculations to compute the third-order bulk modulus $B'$ and inner elastic constants $B_{1xx}$ and $B_{1yy}$. The calculation of $B'$ follows the procedure described in Ref.~\onlinecite{Tanner_PRM_2019} to compute $C_{111}$, $C_{112}$ and $C_{123}$, while $B_{1xx}$ and $B_{1yy}$ were respectively calculated via (mixed) partial derivatives the force $\textbf{F}$ induced by relative displacement of the two atoms in the primitive unit cell with respect to macroscopic (Lagrangian) and internal strains as

\begin{eqnarray}
    B_{1xx} &=&  \frac{ \partial^{2} F_{x} }{ \partial \eta_{1} \partial t_{x} } \bigg|_{ \mathbf{\eta} = \textbf{t} = 0 } \, , \label{eq:B1xx} \\
    B_{1yy} &=& \frac{ \partial^{2} F_{y} }{ \partial \eta_{1} \partial t_{y} } \bigg|_{ \mathbf{\eta} = \textbf{t} = 0 } \, , \label{eq:B1yy}
\end{eqnarray}

\noindent
where $F_{x}$ and $F_{y}$ are respectively the DFT-calculated forces on the displaced atom in the $x$ and $y$ directions, assuming a Cartesian coordinate system whose axes align with the principal cubic crystal axes. Equations~\eqref{eq:B1xx} and~\eqref{eq:B1yy} were evaluated by fitting to forces computed for primitive unit cells to which a Lagrangian strain $\mathbf{\eta} = ( \eta_{1}, 0, 0, 0, 0, 0 )$ was applied, on top of which an internal displacement -- $\textbf{t} = ( t_{x}, 0, 0 )$ for the calculation of $B_{1xx}$, or $\textbf{t} =( 0, t_{y}, 0 )$ for the calculation of $B_{1yy}$ -- was imposed. Here, $\eta_{1}$ was varied between $\pm 0.02$ in steps of $0.01$, while $t_{x}$ and $t_{y}$ were varied from $\pm 6$\% of the equilibrium atomic positions in steps of $3$\%.

The DFT calculations were performed using the projector-augmented wave method, as implemented in the Vienna Ab-initio Simulation Package (VASP). Full details of the DFT calculations -- e.g.~choice of pseudopotentials, cut-off energies, \textbf{k}-point grids, etc.~-- are as described for the LDA calculations of the harmonic elastic properties in Ref.~\onlinecite{Tanner_VFF_2021}. The results of our DFT calculations are summarised in Table~\ref{tab:elastic_parameters}.


\subsection{Anharmonic valence force field potential}
\label{sec:theory_vff}

VFF potentials are appealing due to their combination of physical transparency and parametric simplicity, as well as their ability to be applied to large systems without incurring prohibitive computational expense. Recently, we introduced a fully analytic harmonic VFF potential for the relaxation of diamond (non-polar) group-IV and zinc blende (polar) III-V materials and their alloys. \cite{Tanner_PRB_2019,Tanner_VFF_2021} Here, we are interested solely in the evolution of Raman-active optical phonon modes which, due to the requirement to converse crystal momentum, participate in Raman scattering only when they possess wave vectors in the immediate vicinity of the Brillouin zone centre ($\Gamma$-point). \cite{Parker_PR_1967} We therefore require that the harmonic terms of our VFF potential accurately describe the dispersion of the TO and LO branches close to $\Gamma$, while being less concerned with the accuracy with which the dispersion of the transverse and longitudinal acoustic (TA and LA) branches are described. In an alloy, chemical inhomogeneity (short-range disorder) results in local lattice relaxation that impacts bond vibrational frequencies. To account for this behaviour requires that the strain dependence of the optical phonon frequencies is well described, mandating an anharmonic treatment.

We begin with the harmonic VFF potential of Ref.~\onlinecite{Tanner_VFF_2021} and add three anharmonic terms, which are sufficient to allow for a fully analytic description of the anharmonic bulk modulus $B'$ and the anharmonic inner elastic constants $B_{1xx}$ and $B_{1yy}$. Recalling Eqs.~\eqref{eq:mode_gruneisen_parameter} and~\eqref{eq:phonon_deformation_potential_axial}, we note that an exact reproduction of $B_{xx}$, $B_{1xx}$ and $B_{1yy}$ by an interatomic potential is sufficient to exactly describe the dependence of the optical phonon frequency on hydrostatic and axial strains. We note that this is similar in spirit to the approach -- based on an anharmonic extension of the Keating potential \cite{Rucker_PRB_1995} -- employed by R\"{u}cker and Methfessel in Ref.~\onlinecite{Rucker_PRB_1996} to analyse the vibrational properties of dilute carbide Si$_{1-x}$C$_{x}$ alloys. However, we emphasise that the VFF potential employed here is analytically parametrised, and hence \textit{exactly} reproduces the (selected) harmonic and anharmonic elastic and vibrational properties, and does so while circumventing the conventional requirement to undertake numerical fitting of the VFF force constants.


\begin{table*}[t!]
	\caption{\label{tab:vff_parameters} Equilibrium nearest-neighbour bond length $r_{0}$, harmonic force constants $k_{r}$, $k_{\theta}$, $k_{rr}$ and $k_{r\theta}$, and anharmonic force constants $k_{rrr}$, $k_{\theta\theta\theta}$ and $k_{rr\theta}$ for the VFF potential of Eq.~\eqref{eq:vff_potential}. The force constants were computed using the DFT-calculated structural and elastic properties of Table~\ref{tab:elastic_parameters} in Eqs.~\eqref{eq:force_constant_kr}--~\eqref{eq:force_constant_krrt}.}
    	\begin{ruledtabular}
    		\begin{tabular}{ccccccccc}
    			& $r_{0}$ & $k_{r}$ & $k_{\theta}$ & $k_{rr}$ & $k_{r\theta}$ & $k_{rrr}$ & $k_{\theta\theta\theta}$ & $k_{rr\theta}$ \\
    			& (\AA) & (eV \AA$^{-2}$)  & (eV \AA$^{-2}$ rad$^{-2}$) & (eV \AA$^{-2}$) & (eV \AA$^{-2}$ rad$^{-1}$) & (eV \AA$^{-3}$) & (eV \AA$^{-3}$ rad$^{-3}$) & (eV \AA$^{-3}$ rad$^{-1}$)  \\
    			\hline
    			Ge       & 2.4452  & 7.548763 & 0.424119 & 0.012407 & 0.212248 & $-$36.272488 & 0.410165 & $-$0.779110 \\
       $\alpha$-Sn       & 2.8060  & 5.383901 & 0.219019 & 0.013947 & 0.108082 & $-$22.778825 & 0.195883 & $-$0.376664 \\
    		 zb-GeSn     & 2.6293  & 6.331495 & 0.291446 & 0.014838 & 0.161395 & $-$28.485099 & 0.283400 & $-$0.528379 \\
    		\end{tabular}
    	\end{ruledtabular}
\end{table*}

We consider a non-polar VFF potential in which the lattice free energy per atom $i$, for a single compound, is given by

\begin{widetext}
    \begin{eqnarray}
    	V_{i} &=& \frac{1}{2} \sum_{j} \left[ \frac{ k_{r} }{2} \left( r_{ij} - r_{0} \right)^{2} + \frac{k_{rrr}}{6} \left( r_{ij} - r_{0} \right)^{3} \right] + \sum_{j} \sum_{k > j} \bigg[ \frac{ k_{\theta} }{2} r_{0}^{2} \left( \theta_{ijk} - \theta_{0,ijk} \right)^{2} \nonumber \\
    	&+& \frac{ k_{\theta\theta\theta} }{6} r_{0}^{3} \left( \theta_{ijk} - \theta_{0,ijk} \right)^{3} + k_{r\theta} \bigg( r_{0} \left( r_{ij} - r_{0,ij} \right) + r_{0} \left( r_{ik} - r_{0,ik} \right) \bigg) \left( \theta_{ijk} - \theta_{0,ijk} \right) \nonumber \\
    	&+& k_{rr} \left( r_{ij} - r_{0,ij} \right) \left( r_{ik} - r_{0,ik} \right) + k_{rr\theta} r_{0} \left( r_{ij} - r_{0,ij} \right) \left( r_{ik} - r_{0,ik} \right) \left( \theta_{ijk} - \theta_{0,ijk} \right) \bigg] \, ,
    	\label{eq:vff_potential}
    \end{eqnarray}
\end{widetext}

\noindent
where $j$ and $k$ index the nearest-neighbours of atom $i$, $r_{0,ij}$ and $r_{ij}$ are respectively the equilibrium and relaxed bond lengths between atoms $i$ and $j$, and $\theta_{0,ijk}$ and $\theta_{ijk}$ are respectively the equilibrium and relaxed angles between the nearest-neighbour bonds formed by atoms $i$ and $j$, and atoms $i$ and $k$. The harmonic force constants $k_{r}$ and $k_{\theta}$ respectively describe pure bond stretching and bond-angle bending, while the harmonic force constants $k_{rr}$ and $k_{r\theta}$ represent ``cross terms'' which respectively describe the impact of changes in $r_{ik}$ on $r_{ij}$, and the impact of changes in $\theta_{ijk}$ on $r_{ij}$ and $r_{ik}$. Similarly, the anharmonic force constants $k_{rrr}$ and $k_{\theta\theta\theta}$ respectively describe pure bond stretching and bond-angle bending, while the anharmonic force constant $k_{rr\theta}$ represents a cross-term describing the impact of changes in $\theta_{ijk}$ on $r_{ij}$ and $r_{ik}$.

To apply Eq.~\eqref{eq:vff_potential} to a Ge$_{1-x}$Sn$_{x}$ alloy supercell we utilise a simple arithmetic averaging of the three-body force constants. \cite{Lopuszynski_JPCM_2010,Mattila_JAP_1999,Branicio_APL_2003} Here, e.g., the product $k_{\theta} r_{0}^{2}$ of the bond-angle bending force constant $k_{\theta}$ and (squared) equilibrium nearest-neighbour bond length $r_{0}$ for a Ge-Ge-Sn triplet -- i.e.~a pair of Ge-Ge and Ge-Sn nearest-neighbour bonds sharing a common Ge atom -- is averaged as

\begin{equation}
    k_{\theta} r_{0}^{2} \longrightarrow \frac{1}{2} \bigg( \underbrace{ k_{\theta} ( \text{Ge} ) r_{0}^{2} ( \text{Ge} ) }_{ \text{Ge-Ge bond} } + \underbrace{ k_{\theta} ( \text{GeSn} ) r_{0}^{2} ( \text{GeSn} ) }_{ \text{Ge-Sn bond} } \bigg) \, .
\end{equation}

All other three-body force constants are averaged similarly, except for the $k_{r\theta}$ cross-term force constant, where we employ a mixing of the force constants associated with the compound materials formed by the two atoms constituting each nearest-neighbour bond giving, for the Ge-Ge-Sn triplet

\begin{widetext}
    \begin{equation}
        k_{r\theta} \bigg( r_{0} \left( r_{ij} - r_{0,ij} \right) + r_{0} \left( r_{ik} - r_{0,ik} \right) \bigg) \longrightarrow \underbrace{ k_{r\theta} ( \text{Ge} ) r_{0} ( \text{Ge} ) \left( r_{ij} - r_{0} ( \text{Ge} ) \right) }_{ \text{Ge-Ge bond} } + \underbrace{ k_{r\theta} ( \text{GeSn} ) r_{0} ( \text{GeSn} ) ( r_{ik} - r_{0} ( \text{GeSn} ) ) }_{ \text{Ge-Sn bond} } \, .
    \end{equation}
\end{widetext}

In Ref.~\onlinecite{Tanner_PRB_2019} we showed that it is possible, by expanding Eq.~\eqref{eq:vff_potential} to second order in macroscopic and internal strains, to obtain analytical expressions linking the linear elastic properties to the harmonic force constants. These expressions can then be inverted to obtain analytical expressions for the four independent harmonic force constants in terms of a choice of four independent harmonic elastic properties. In Ref.~\onlinecite{Tanner_VFF_2021} we chose to parametrise the harmonic VFF potential via analytical expressions linking the relaxed second-order elastic constants $C_{11}$, $C_{12}$ and $C_{44}$, and the Kleinman parameter $\zeta$, to the harmonic force constants $k_{r}$, $k_{\theta}$, $k_{rr}$ and $k_{r\theta}$, and in so doing obtained an exact description of the \textit{static} properties of the lattice in the linear elastic limit. Here, we are concerned with \textit{dynamic} lattice properties, so revise the parametrisation of the harmonic terms to omit $\zeta$ in favour of the inner elastic constant $B_{xx}$, and in so doing obtain an exact description of the zone-centre optical phonon frequency $\omega_{\scalebox{0.7}{\text{TO}}}$ (cf.~Eq.~\eqref{eq:optical_phonon_frequency}). The resulting expressions for the harmonic force constants are

\begin{eqnarray}
    k_{r} &=& \frac{ r_{0} }{ \sqrt{ 3 } } \left( 5 C_{11} - 2 C_{12} + 3 B_{xx} r_{0}^{2} \right) \, \nonumber \\
    &-& 4 r_{0}^{2} \sqrt{ \left( C_{11} - C_{12} - C_{44} \right) B_{xx} } \, , \label{eq:force_constant_kr} \\
    k_{\theta} &=& \frac{ 2 r_{0} }{ 3 \sqrt{3} } \left( C_{11} - C_{12} \right) \, , \label{eq:force_constant_kt} \\
    k_{rr} &=& - \frac{ r_{0} }{ 6 \sqrt{3} } \left( C_{11} - 10 C_{12} + 3 B_{xx} r_{0}^{2} \right)  \, \nonumber \\
    &+& \frac{ 2 r_{0}^{2} }{ 3 } \sqrt{ \left( C_{11} - C_{12} - C_{44} \right) B_{xx} } \, , \label{eq:force_constant_krr} \\
    k_{r\theta} &=& \frac{ 2 r_{0} }{ 3 } \sqrt{ \frac{ 2 }{ 3 } } \left( C_{11} - C_{12} \right)  \, \nonumber \\
    &-& \frac{ \sqrt{2} r_{0}^{2} }{ 3 } \sqrt{ \left( C_{11} - C_{12} - C_{44} \right) B_{xx} } \, , \label{eq:force_constant_krt}
\end{eqnarray}

\noindent
which can be evaluated using our DFT-calculated equilibrium bond lengths $r_{0}$ ($= \frac{ \sqrt{3} }{4} \, a_{0}$), relaxed elastic constants $C_{11}$, $C_{12}$ and $C_{44}$, and inner elastic constant $B_{xx}$ (cf.~Table~\ref{tab:elastic_parameters}).

Similarly, by expanding Eq.~\eqref{eq:vff_potential} to third order in macroscopic and internal strains, we obtain the following expressions for the anharmonic force constants

%

\begin{eqnarray}
    k_{rrr} &=& 12 \, \sqrt{3} \, B' \, , \label{eq:force_constant_krrr} \\
    k_{\theta\theta\theta} &=& - \frac{ 1 }{ 2 \sqrt{6} } \left( 2 C_{11} + C_{12} + \frac{9}{2} \, B' \right)  \, \nonumber \\
    &-& \frac{ r_{0}^{2} }{ 4 \sqrt{6} } \left( B_{1xx} - 4 B_{1yy} - 4 B_{xx} \right)  \, \nonumber \\
    &-& \frac{ 7 r_{0} }{ 6 \sqrt{2} } \sqrt{ \left( C_{11} - C_{12} - C_{44} \right) B_{xx} } \, , \label{eq:force_constant_kttt} \\
    k_{rr\theta} &=& - \frac{ 1 }{ \sqrt{6} } \left( C_{11} - 4 C_{12} - \frac{9}{2} \, B' \right)  \, \nonumber \\
    &-& \frac{ r_{0}^{2} }{ 2 \sqrt{6} } \left( B_{1xx} + 2 B_{1yy} \right)  \, \nonumber \\
    &+& \frac{ r_{0} }{ 3 \sqrt{2} } \sqrt{ \left( C_{11} - C_{12} - C_{44} \right) B_{xx} } \, , \label{eq:force_constant_krrt}
\end{eqnarray}

\noindent
which we note are determined by a combination of the second- and third-order elastic constants.

Eqs.~\eqref{eq:force_constant_kr} --~\eqref{eq:force_constant_krrt} allow to directly compute the
force constants of Eq.~\eqref{eq:vff_potential} for each of Ge, $\alpha$-Sn and zb-GeSn. The VFF force constants computed in this manner, using the data of Table~\ref{tab:elastic_parameters}, are listed in Table~\ref{tab:vff_parameters}. We emphasise that the resulting VFF potentials for Ge, $\alpha$-Sn and zb-GeSn \textit{exactly} reproduce \textit{all} of the quantities listed in Table~\ref{tab:elastic_parameters}, representing a significant improvement in accuracy over anharmonic VFF potentials employed in previous analyses of Raman scattering in group-IV alloys. While we restrict our attention to Ge$_{1-x}$Sn$_{x}$ alloys here, we note that Eqs.~\eqref{eq:force_constant_kr} --~\eqref{eq:force_constant_krrt} apply generally, allowing for a fully analytic parametrisation of Eq.~\eqref{eq:vff_potential} for any diamond-structured material for which the structural and elastic properties listed in Table~\ref{tab:elastic_parameters} are known.


\begin{figure*}[t!]
	\includegraphics[width=1.00\textwidth]{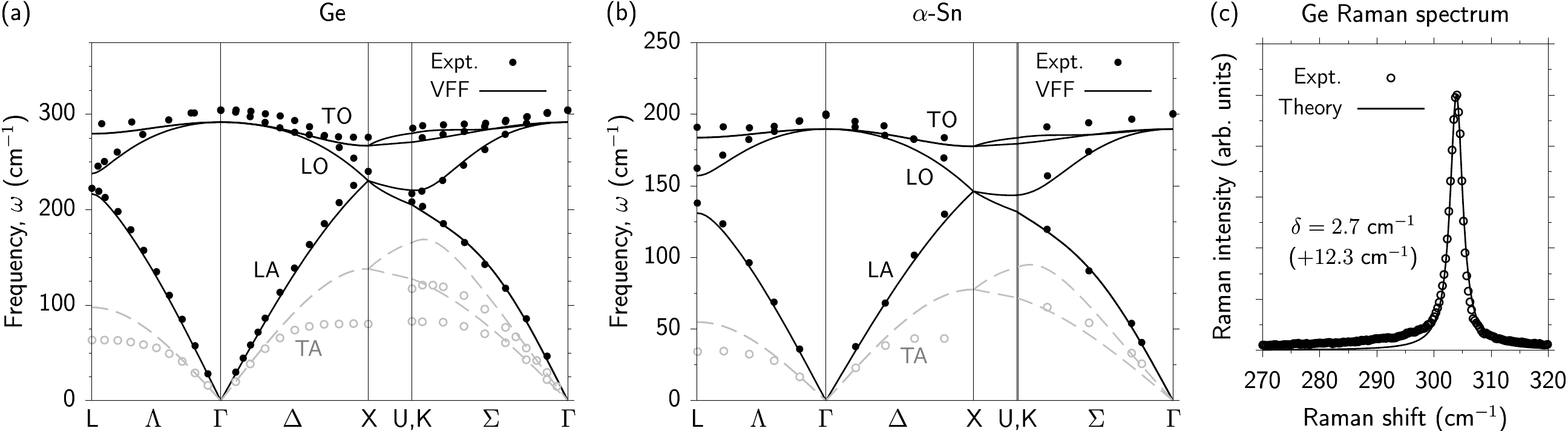}
	\caption{Calculated phonon band structure (solid black and dashed grey lines) of (a) Ge, and (b) $\alpha$-Sn, using the anharmonic VFF potential of Eq.~\eqref{eq:vff_potential} and Table~\ref{tab:vff_parameters}. Experimental data (closed black and open grey circles) for Ge and $\alpha$-Sn are from Refs.~\onlinecite{Nilsson_PRB_1971} and~\onlinecite{Price_PRB_1971}, respectively. The VFF potential fails to describe the TA phonon dispersion away from the zone centre (dashed grey lines vs.~open grey circles), but describes well the dispersion of the LA, LO and TO branches (solid black lines vs.~closed black circles). (c) Ge-Ge mode Raman shift for a pure Ge NW, calculated via Eq.~\eqref{eq:raman_tensor} assuming a frequency linewidth $\delta = 2.7$ cm$^{-1}$ (solid black lines), and measured as described in Sec.~\ref{sec:experiment_raman} (open black circles). The calculated Raman spectrum has been blueshifted by 12.3 cm$^{-1}$ to account for the underestimation of $\omega_{\scalebox{0.7}{\text{TO}}}$ ($= \omega_{\scalebox{0.7}{\text{LO}}}$) by LDA-DFT.}
	\label{fig:benchmark_calculations}
\end{figure*}

The phonon band structures of Ge and $\alpha$-Sn calculated using Eq.~\eqref{eq:vff_potential}, parametrised via Table~\ref{tab:vff_parameters}, are shown in Figs.~\ref{fig:benchmark_calculations}(a) and~\ref{fig:benchmark_calculations}(b), respectively. The closed black and open grey circles in Figs.~\ref{fig:benchmark_calculations}(a) and~\ref{fig:benchmark_calculations}(b) respectively denote the phonon bands obtained from the inelastic neutron scattering measurements of Refs.~\onlinecite{Nilsson_PRB_1971} and~\onlinecite{Price_PRB_1971}. For both Ge and $\alpha$-Sn we note a typical VFF description of the phonon bands: the potential describes well the measured dispersion of the LA, LO and TO bands (solid black lines vs.~closed black circles), but fails to capture accurately the full dispersion of the TA bands, tending to overestimate the TA phonon frequency by failing to capture the softening of these modes towards the zone boundaries (dashed grey lines vs.~open grey circles). \cite{Tanner_PRB_2019} We emphasise that the VFF potential exactly reproduces the relaxed second-order elastic constants $C_{ij}$, so that the zone-centre LA and TA phonon group velocities are correctly described. \cite{Tanner_VFF_2021} We also note that it is in principle possible to soften the dispersion of the TA bands and bring them into quantitative agreement with experiment, by extending the VFF potential to incorporate angular interactions involving four co-planar bonds. \cite{Cousins_PRB_2003,Paul_JCE_2010,Steiger_PRB_2011,Barrett_PRB_2015} However, we do not pursue such an approach in this work since we are interested solely in near-zone-centre Raman-active optical phonon modes, and including such additional terms in our VFF potential would significantly complicate our analytical parametrisation without impacting the calculated Raman spectra. Finally, we note that the underestimation of the measured optical phonon frequencies in Figs.~\ref{fig:benchmark_calculations}(a) and~\ref{fig:benchmark_calculations}(b) is not an intrinsic failure of the VFF potential, but rather a consequence of the slight underestimation of $\omega_{\scalebox{0.7}{\text{TO}}}$ ($= \omega_{\scalebox{0.7}{\text{LO}}}$) by LDA-DFT, \cite{Tanner_VFF_2021} with these LDA-DFT calculated values having been employed in the analytical evaluation of the VFF force constants. A small, rigid upward shift of the calculated LO and TO bands -- based, e.g., on using parameters obtained from hybrid functional DFT calculations \cite{Tanner_VFF_2021} -- would bring their dispersion into quantitative agreement with experiment.


\subsection{Calculation of Raman spectra}
\label{sec:theory_raman}

For a given Ge$_{1-x}$Sn$_{x}$ alloy supercell, we firstly employ the anharmonic VFF potential of Eq.~\eqref{eq:vff_potential} to perform lattice relaxation via minimisation of the lattice free energy. Next, the VFF potential is used to compute the supercell dynamical matrix, which is diagonalised to obtain the  (normal) phonon modes -- i.e.~the phonon frequencies $\omega_{\nu}$ and displacements $\textbf{u}_{\nu}$. All VFF calculations were performed by implementing Eq.~\eqref{eq:vff_potential} using the General Utility Lattice Program (GULP). \cite{Gale_JCSFT_1997,Gale_MS_2003,Gale_ZK_2005}

Using the computed phonon modes for a given supercell, the Raman spectrum -- i.e.~the frequency-dependent Raman polarisability tensor -- is computed as \cite{Baroni_PRL_1990,Rucker_PRB_1996}

\begin{equation}
    \sigma_{\alpha\beta} ( \omega ) \propto \sum_{\nu} \frac{ 2 \omega \delta }{ \left( \omega^{2} - \omega_{\nu}^{2} \right)^{2} + \omega^{2} \delta^{2} } \left| \sum_{i} s_{i} \, \epsilon_{\alpha \beta \gamma} \, \vert \widehat{\textbf{$\gamma$}} \cdot \textbf{u}_{\nu,i} \vert^{2} \, \right|^{2}
    \label{eq:raman_tensor}
\end{equation}

\noindent
where $\delta$ is the frequency linewidth, the outer sum runs over supercell zone-centre phonon modes $\nu$ having frequencies $\omega_{v}$, the inner sum runs over atoms $i$, with $\widehat{\textbf{$\gamma$}} \cdot \textbf{u}_{\nu,i}$ being the displacement of atom $i$ along the Cartesian direction $\gamma = x, y$ or $z$ in mode $\nu$ (the Raman tensor indices $\alpha$ and $\beta$ also represent Cartesian directions), $\epsilon_{\alpha \beta \gamma}$ is the Levi-Civita symbol, and $s_{i} = \pm 1$ depending on which of the two face-centred cubic sublattices atom $i$ lies. For a given alloy supercell we obtain the Raman spectrum $\sigma ( \omega )$ by averaging over the Raman tensor components of Eq.~\eqref{eq:raman_tensor} as $\sigma ( \omega ) = \frac{1}{3} ( \sigma_{xy} ( \omega ) + \sigma_{xz} ( \omega ) + \sigma_{yz} ( \omega ) )$.

Figure~\ref{fig:benchmark_calculations}(c) compares our calculated (solid black lines) and measured (open black circles) Raman spectra for a pure Ge NW. Here, as the NW diameter is sufficiently large to preclude the formation of bound phonon modes, we perform the theoretical calculation for a bulk-like supercell (i.e.~employing Born-von Karman boundary conditions). We retain this approach for our Ge$_{1-x}$Sn$_{x}$ alloy calculations in Sec.~\ref{sec:results}, below. Our measured optical phonon frequency -- i.e.~the peak corresponding to the observed Ge-Ge mode -- of 303.9 cm$^{-1}$ is 12.3 cm$^{-1}$ larger than the value $\omega_{\scalebox{0.7}{\text{TO}}} = 291.6$ cm$^{-1}$ obtained from our LDA-DFT calculations and employed in our VFF parametrisation (cf.~Table~\ref{tab:elastic_parameters}). To compare the calculated and measured spectra we therefore rigidly blueshift the calculated spectrum by 12.3 cm$^{-1}$. We then treat the linewidth $\delta$ in Eq.~\eqref{eq:raman_tensor} as an empirical parameter, which we adjust to match the spectral width of the measured Raman spectrum. With an empirically parametrised Ge-Ge mode linewidth $\delta = 2.7$ cm$^{-1}$ we note that the calculated Raman spectrum describes well our measured spectrum, confirming that the VFF potential provides a suitable platform to undertake direct atomistic calculations and analysis of Raman spectra.


\section{Results}
\label{sec:results}


\subsection{Sn composition-dependent Raman shift}
\label{sec:results_composition}


\begin{figure*}[t!]
	\includegraphics[width=1.00\textwidth]{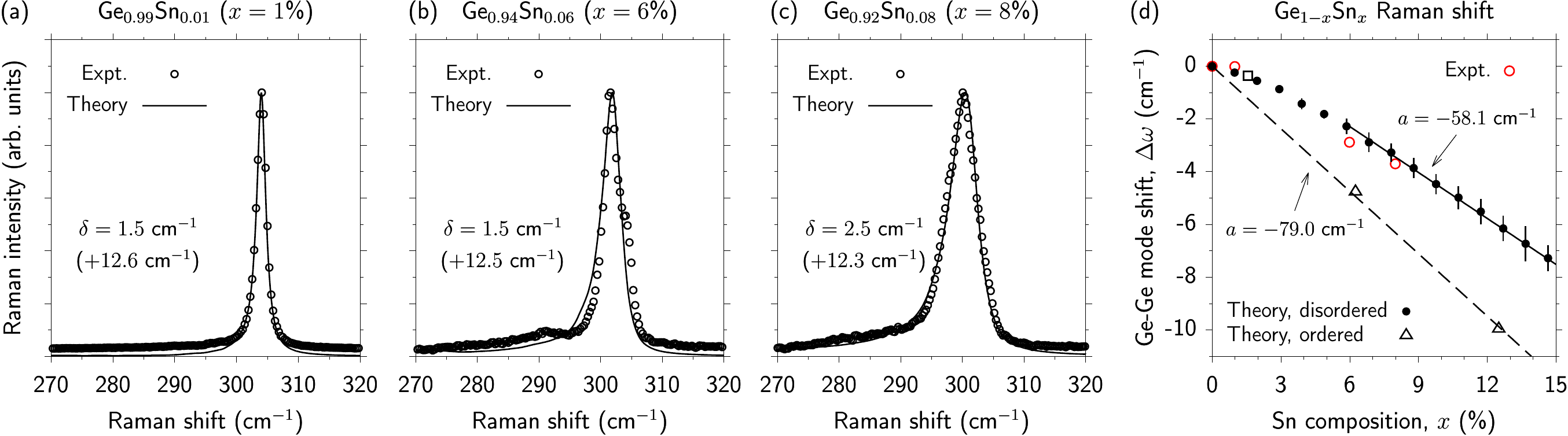}
	\caption{Measured (open circles) and calculated (solid lines) Raman spectra for Ge$_{1-x}$Sn$_{x}$ NWs having Sn composition (a) $x = 1$\%, (b) $x = 6$\%, and (c) $x = 8$\%. At each composition the calculated Raman spectrum was obtained by averaging over the spectra calculated for 25 distinct disordered, relaxed 512-atom Ge$_{1-x}$Sn$_{x}$ alloy supercells, and rigidly blueshifted by the amount listed in parentheses to align with the peak of the measured spectrum. The best-fit linewidth $\delta$ is listed for the calculated Raman spectra in (a), (b) and (c) (cf.~Eq.~\eqref{eq:raman_tensor}). (d) Measured (open red circles) and calculated (closed black circles) shift $\Delta \omega$ of the Ge-Ge mode frequency as a function of Sn composition $x$. The theoretical data points represent the average value of $\Delta \omega$ calculated for the 25 distinct disordered alloy supercells considered at each Sn composition; error bars represent the associated standard deviation of the calculated $\Delta \omega$ values for the 25 supercells. The open black square shows the calculated $\Delta \omega$ for an ordered 64-atom Ge$_{63}$Sn$_{1}$ ($x = 1.56$\%) supercell, while the open black triangles show the calculated $\Delta \omega$ for ordered 16-atom Ge$_{15}$Sn$_{1}$ ($x = 6.25$\%) and 8-atom Ge$_{7}$Sn$_{1}$ ($x = 12.5$\%) supercells. The solid black line is a linear fit to the calculated disordered supercell $\Delta \omega$ for $x \geq 6$\%; the dashed black line is a linear fit to the calculated $\Delta \omega$ for the ordered Ge$_{15}$Sn$_{1}$ and Ge$_{7}$Sn$_{1}$ supercells.}
	\label{fig:GeSn_theory_vs_experiment}
\end{figure*}


Our measured Raman spectra for the $x = 1$, 6 and 8\% Ge$_{1-x}$Sn$_{x}$ NWs are shown using open black circles in Figs.~\ref{fig:GeSn_theory_vs_experiment}(a), \ref{fig:GeSn_theory_vs_experiment}(b) and \ref{fig:GeSn_theory_vs_experiment}(c), respectively. The evolution of the measured Raman spectra is in line with that described elsewhere in the literature for Ge$_{1-x}$Sn$_{x}$: with increasing Sn composition $x$ we observe a composition-dependent redshift $\Delta \omega$ of the Raman-active Ge-Ge optical mode frequency, in addition to spectral broadening of the Ge-Ge mode Raman peak. The corresponding measured frequency shift $\Delta \omega$ of the Ge-Ge optical mode as a function of $x$ is shown in Fig.~\ref{fig:GeSn_theory_vs_experiment}(d) using open red circles. Typically, this frequency shift is assumed to be linear in the composition $x$ in a relaxed alloy, and to have a linear dependence on in-plane strain $\epsilon_{\parallel}$ in a pseudomorphically strained epitaxial layer, such that \cite{Cerdeira_PRB_1972,Rojas-Lopez_JAP_1998,Lin_APL_2011,Su_SSC_2011}

\begin{equation}
    \Delta \omega ( x, \epsilon ) \equiv \Delta \omega ( x ) + \Delta \omega ( \epsilon ) = a \, x + b \, \epsilon_{\parallel} \, .
    \label{eq:mode_frequency_shift}
\end{equation}

\noindent
where $\Delta \omega_{x}$ and $\Delta \omega ( \epsilon )$ are, respectively, the composition- and strain-induced contributions to the mode frequency shift, $\epsilon_{\parallel}$ is the lattice mismatch in the growth plane, and $a$ and $b$ are the so-called composition and strain shift coefficients.

Our measurements are performed on free-standing and fully relaxed Ge$_{1-x}$Sn$_{x}$ NWs, so that there is no strain-related contribution to our measured values of $\Delta \omega$. Performing a forced zero-intercept linear fit to the experimental data depicted by open red circles in Fig.~\ref{fig:GeSn_theory_vs_experiment}(d) yields $a = -46.5$ cm$^{-1}$. We note that this value of $a$ is lower in magnitude than the majority of values reported in the literature -- see, e.g., Table I of Ref.~\onlinecite{Gassenq_APL_2017}, for a summary of these experimentally extracted composition shift coefficients -- which are typically in the range of $-60$ to $-100$ cm$^{-1}$. We note however that the determination of $a$ in the majority of previous studies relies on correction of $\Delta \omega$ measured for pseudomorphically strained or partially relaxed epitaxial Ge$_{1-x}$Sn$_{x}$ layers or etched Ge$_{1-x}$Sn$_{x}$ nanostructures to compensate for the impact of strain, a procedure which has the potential to introduce significant parametric uncertainty. \cite{Li_APL_2004}


In order to analyse theoretically the evolution of the Raman spectrum in Ge$_{1-x}$Sn$_{x}$ with $x$ we perform high-throughput calculations for a series of relaxed 512-atom ($4 \times 4 \times 4$ simple cubic) alloy supercells. At each Sn composition considered we construct 25 distinct randomly disordered Ge$_{512-N}$Sn$_{N}$ ($x = \frac{N}{512}$) supercells, and for each individual supercell we use the VFF potential of Sec.~\ref{sec:theory_vff} to perform lattice relaxation and compute the Raman spectrum following the procedure outlined in Sec.~\ref{sec:theory_raman}. We then perform an (unweighted) averaging of these 25 Raman spectra to obtain a single Raman spectrum for the chosen Sn composition to compare to experiment. The results of these calculations at $x = 1$, 6 and 8\% are shown in Figs.~\ref{fig:GeSn_theory_vs_experiment}(a), \ref{fig:GeSn_theory_vs_experiment}(b) and \ref{fig:GeSn_theory_vs_experiment}(c) using solid black lines. In each case, the calculated Raman spectrum has been (i) normalised to the measured Raman spectrum at the peak value, and (ii) rigidly blueshifted to align with the measured peak frequency, compensating for the underestimation of optical phonon frequencies by the LDA-DFT calculations used to parametrise the VFF potential (cf.~Sec.~\ref{sec:theory_dft}).

Recalling that our LDA-DFT calculations, and hence VFF phonon mode calculations, underestimate $\omega_{\scalebox{0.7}{\text{TO}}}$ in Ge by 12.3 cm$^{-1}$ compared to experiment, we note that the blueshifts employed to align the calculated and measured Raman peaks -- listed in parentheses in Figs.~\ref{fig:GeSn_theory_vs_experiment}(a), \ref{fig:GeSn_theory_vs_experiment}(b) and \ref{fig:GeSn_theory_vs_experiment}(c) -- are within 0.3 cm$^{-1}$ of this value, suggesting that our theoretical calculations describe accurately the Sn-induced redshift of the Ge-Ge mode frequency observed in our experimental measurements. At each Sn composition $x$ considered in our theoretical calculations we directly extract $\Delta \omega$ as the difference between the frequency at which the calculated Raman spectral peak is located and the calculated value of $\omega_{\scalebox{0.7}{\text{TO}}}$ ($= \omega_{\scalebox{0.7}{\text{LO}}}$) in pure Ge. The results of these calculations are summarised in Fig.~\ref{fig:GeSn_theory_vs_experiment}(d), where closed black circles show the configuration-averaged value of $\Delta \omega$, computed at each Sn composition as the average of the values of $\Delta \omega$ calculated for the 25 distinct disordered supercells considered at that composition. The associated error bars denote the standard deviation of the $\Delta \omega$ values computed for the 25 distinct disordered supercells considered at each Sn composition. We note excellent quantitative agreement between our randomly disordered relaxed alloy supercell calculations and relaxed NW experimental measurements. Additionally, we note that our configuration-averaged theoretical calculations predict bowing of the Ge-Ge mode frequency in the composition range $x \lesssim 6$\%, contrary to the usual assumption of linear variation with $x$. As such, we do not attempt a linear fit to our calculated $\Delta \omega (x)$ data across the full composition range considered. However, we note that for $x \gtrsim 6$\% the calculated $\Delta \omega (x)$ is approximately linear in $x$. In the $x = 6$ -- 15\% composition range a linear fit to our theoretical data yields $a = -58.1$ cm$^{-1}$ (denoted by a solid black line in Fig.~\ref{fig:GeSn_theory_vs_experiment}(d)), which is at the low end of the range of values reported in the experimental literature. \cite{Rojas-Lopez_JAP_1998,Li_APL_2004,Costa_SSC_2007,Lin_APL_2011,Su_SSC_2011,Fournier-Lupien_APL_2013,Cheng_ECSJSSST_2013,Chang_TSF_2015,Takeuchi_JJAP_2016,Gassenq_APL_2017,Xu_APL_2019,Bouthillier_SST_2020} Test calculations for larger supercells containing up to 1728 atoms (not shown) yield minimal changes in $\Delta \omega (x)$, confirming that the 512-atom supercells employed in our analysis are sufficiently large to allow for realistic analysis of disorder-related effects (since, in smaller supercells, Born-von Karman boundary conditions can introduce spurious short-range ordering which limits the ability to describe realistic disordered alloy microstructure \cite{Broderick_GePb_2019}).


\subsection{Impact of alloy disorder}
\label{sec:results_disorder}


Also shown in Fig.~\ref{fig:GeSn_theory_vs_experiment}(d) are the calculated values of $\Delta \omega$ for three relaxed, \textit{ordered} Ge$_{1-x}$Sn$_{x}$ alloy supercells: (i) a $2 \times 2 \times 2$ simple cubic Ge$_{63}$Sn$_{1}$ ($x = 1.56$\%; open black square) supercell, and (ii) a $2 \times 2 \times 2$ face-centred cubic Ge$_{15}$Sn$_{1}$ ($x = 6.25$\%; open black triangle) supercell, and (iii) a $1 \times 1 \times 1$ simple cubic Ge$_{7}$Sn$_{1}$ ($x = 12.5$\%; open black triangle) supercell. We note that the calculated Ge-Ge mode shift $\Delta \omega = 4.8$ cm$^{-1}$ for the ordered Ge$_{15}$Sn$_{1}$ supercell is approximately twice as large in magnitude than the corresponding value $\Delta \omega = 2.4 \pm 0.4$ cm$^{-1}$ obtained via configurational averaging for a series of disordered 512-atom Ge$_{480}$Sn$_{32}$ supercells having the same Sn composition $x = 6.25$\%. Application of a purely linear, forced zero-intercept fit $\Delta \omega (x) = a \, x$ using only these single ordered and disordered values of $\Delta \omega$ yields respective $a$ values of $-76.2$ and $-39.0$ cm$^{-1}$. Our calculations therefore suggest that the presence of short-range alloy disorder in a purely substitutional Ge$_{1-x}$Sn$_{x}$ alloy has a significant impact on the magnitude of $\Delta \omega$ at fixed $x$, with larger shifts occurring in the presence of short-range ordering, a finding which may in part explain the significant spread in $a$ values reported in previous experimental studies.

Indeed, recent systematic measurements by Wang et al. \cite{Wang_IEEETED_2020} demonstrated strong changes in $\Delta \omega$ associated with Ge-Ge optical modes in as-grown vs.~annealed Ge$_{1-x}$Sn$_{x}$. Specifically, the results of Ref.~\onlinecite{Wang_IEEETED_2020} demonstrate an increase (decrease) after annealing in the magnitude of $\Delta \omega$ at Sn composition $x = 3$\% ($x = 10$\%), with the additional complication that the observed change in $\Delta \omega$ is itself a function of the annealing temperature. Similar systematic measurements by Gassenq et al. \cite{Gassenq_APL_2017} demonstrated strong dependence of the magnitude of $\Delta \omega$ associated with Ge-Ge optical modes on Ge$_{1-x}$Sn$_{x}$ epitaxial layer thickness. These experimental results highlight the strong sensitivity of the measured Ge-Ge mode Raman shift to the Ge$_{1-x}$Sn$_{x}$ alloy microstructure and the presence of native defects. This, combined with the aforementioned empirical determination of the composition and strain shift coefficients $a$ and $b$, makes quantitative comparison of our calculated values of $a$ and $b$ -- which are for idealised substitutional, defect-free alloys -- to experiment challenging.

\begin{figure}[t!]
	\includegraphics[width=1.00\columnwidth]{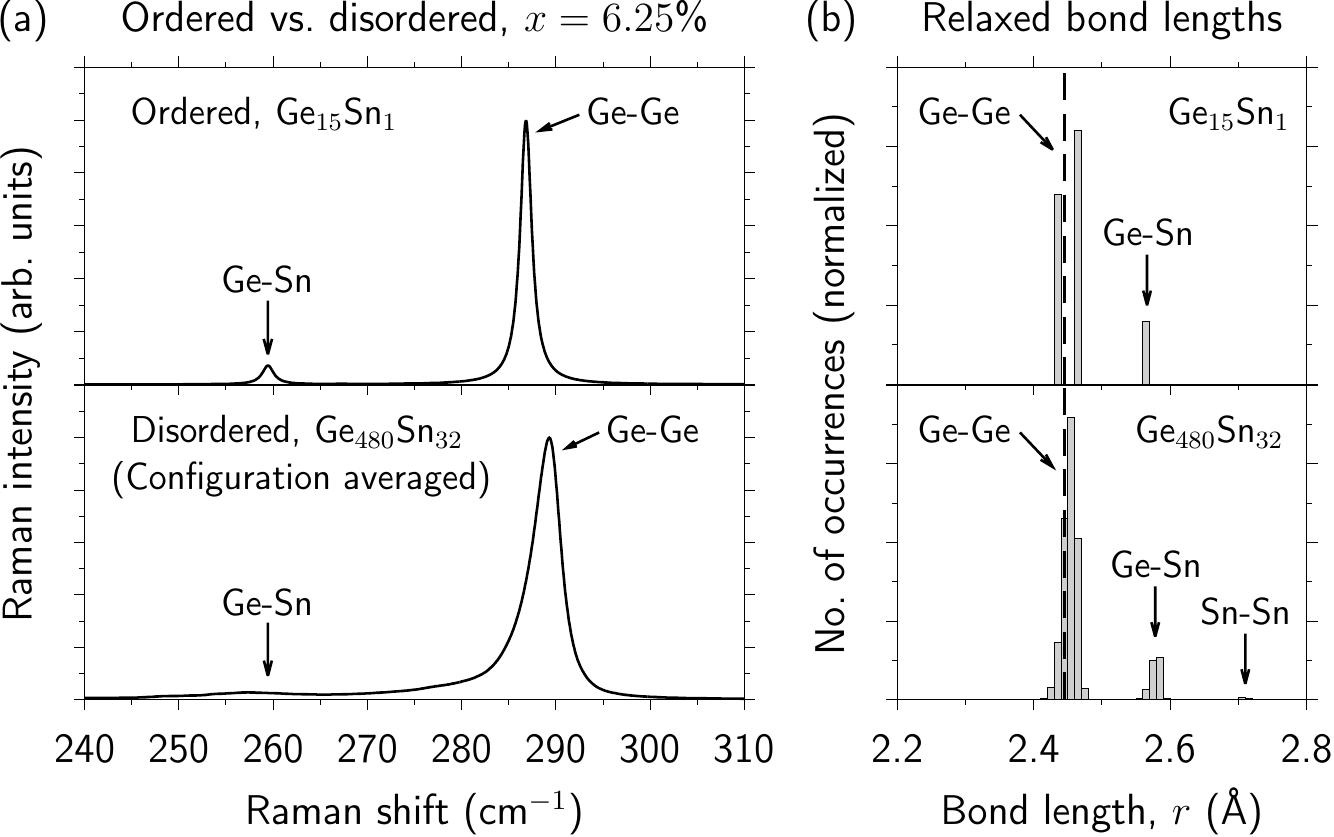}
	\caption{Calculated (a) Raman spectra, and (b) relaxed nearest-neighbour bond length distributions, for ordered (upper panels) and disordered (lower panels) Ge$_{1-x}$Sn$_{x}$ alloy supercells having the same Sn composition $x = 6.25$\%. The lower panels show the configuration-averaged Raman spectrum and bond length distribution calculated for a series of 25 disordered Ge$_{480}$Sn$_{32}$ supercells. The vertical dashed line in (b) denotes the unstrained Ge-Ge bond length $r_{0}$ (cf.~Table~\ref{tab:vff_parameters}). In the disordered case we note the reduced shift in Ge-Ge mode frequency (cf.~Fig.~\ref{fig:GeSn_theory_vs_experiment}(d)), enhanced broadening of the Ge-Ge mode feature on the low-frequency side of the Raman peak, and significant broadening of the Ge-Sn mode feature.}
	\label{fig:GeSn_raman_ordered_vs_disordered}
\end{figure}

To highlight the impact of alloy disorder in a substitutional Ge$_{1-x}$Sn$_{x}$ alloy, in Fig.~\ref{fig:GeSn_raman_ordered_vs_disordered}(a) we compare the calculated Raman spectra of ordered and disordered supercells having the same Sn composition, $x = 6.25$\%: an ordered 16-atom Ge$_{15}$Sn$_{1}$ supercell (upper panel), and configuration averaging over 25 distinct, disordered 512-atom Ge$_{480}$Sn$_{32}$ supercells (lower panel). Figure~\ref{fig:GeSn_raman_ordered_vs_disordered}(b) shows the distribution of relaxed nearest-neighbour bond lengths in these supercells, sorted into bins of width 0.01 \AA, \cite{Halloran_OQE_2019,Tanner_VFF_2021} where the distribution for the disordered Ge$_{480}$Sn$_{32}$ supercells has again been obtained via configurational averaging. Note that in Fig.~\ref{fig:GeSn_raman_ordered_vs_disordered}(a) we have not applied the rigid $\approx 12.3$ cm$^{-1}$ blueshift employed to facilitate comparison to experiment in Figs.~\ref{fig:benchmark_calculations}(c) and Figs.~\ref{fig:GeSn_theory_vs_experiment}(a) --~\ref{fig:GeSn_theory_vs_experiment}(c), but instead focus here on a direct comparison of the as-computed Raman spectra for the ordered and disordered supercells.

For the ordered supercell, we observe two features in the Raman spectrum: the primary Ge-Ge mode peak at a frequency of 286.8 cm$^{-1}$, and a mode peak at a frequency of 259.5 cm$^{-1}$ associated with the four nearest-neighbour Ge-Sn bonds in the supercell. From Table~\ref{tab:elastic_parameters} we recall that the Ge-Sn optical mode frequency is 234.1 cm$^{-1}$. The appearance of the Ge-Sn feature at higher frequency in the ordered Ge$_{15}$Sn$_{1}$ supercell is readily understood in terms of the alloy lattice relaxation. In this ordered supercell, the four Ge-Sn neighbour bonds do not relax to their zb-GeSn equilibrium bond length (cf.~Table~\ref{tab:vff_parameters}) but, with a relaxed length $r = 2.562$ \AA, are compressed by $\approx 2.5$\% due to the pressure exerted by the surrounding Ge lattice. \cite{Tanner_VFF_2021} Converting this bond length to a lattice parameter, this compressed Ge-Sn bond length is equivalent to an $\approx 7.4$\% volume compression in a 2-atom zb-GeSn primitive unit cell. Using this change in volume in conjunction with the computed zb-GeSn zone-centre TO mode Gr\"{u}neisen parameter (cf.~Table~\ref{tab:elastic_parameters}), we estimate the optical vibrational frequency of the strained Ge-Sn bonds via $\frac{ \Delta \omega_{\scalebox{0.5}{\text{TO}}} }{ \omega_{{\scalebox{0.5}{\text{TO}}},0} } = \gamma_{\scalebox{0.7}{\text{TO}}} ( \frac{ \Delta \Omega }{ \Omega_{0} } )$, where the subscript ``0'' denotes unstrained quantities, obtaining a value of 256.1 cm$^{-1}$ for $\omega_{\scalebox{0.6}{\text{TO}}}$. This estimated frequency, despite neglecting changes in Ge-Ge bond lengths at lattice sites surrounding the substitutional Sn impurity, is in close agreement with the observed Ge-Sn mode peak frequency of 259.5 cm$^{-1}$ from the full supercell calculation.

Turning our attention to the configuration-averaged disordered alloy Raman spectrum in the lower panel of Fig.~\ref{fig:GeSn_raman_ordered_vs_disordered}(a), we note three key qualitative differences compared to the ordered supercell Raman spectrum of the upper panel. Firstly, the magnitude of the redshift $\Delta \omega$ of the Ge-Ge mode peak frequency -- i.e.~relative to the zone-centre optical mode frequency $\omega_{\scalebox{0.7}{\text{TO}}} = \omega_{\scalebox{0.7}{\text{LO}}}$ of pure Ge -- is reduced compared to the ordered case. The peak frequency of 286.8 cm$^{-1}$ in the ordered Ge$_{15}$Sn$_{1}$ supercell constitutes a redshift of 4.8 cm$^{-1}$ compared to the optical phonon frequency 291.6 cm$^{-1}$ of pure Ge (cf.~Table~\ref{tab:elastic_parameters}), while the peak frequency of 289.3 cm$^{-1}$ in the configuration-averaged disordered Ge$_{480}$Sn$_{32}$ supercells represents a redshift of 2.3 cm$^{-1}$, which is lower by a factor of $\approx 2$ compared to the ordered case (cf.~Fig.~\ref{fig:GeSn_theory_vs_experiment}(d)).


Secondly, comparing the upper and lower panels of Fig.~\ref{fig:GeSn_raman_ordered_vs_disordered}(a), we note the enhanced and asymmetric nature of the spectral broadening of the Ge-Ge mode Raman peak in the disordered case. For both the ordered and disordered calculations of Fig.~\ref{fig:GeSn_raman_ordered_vs_disordered}(a) we have assumed an equal frequency linewidth $\delta = 1.5$ cm$^{-1}$ in Eq.~\eqref{eq:raman_tensor}. The difference in spectral broadening between the ordered and disordered supercell Raman spectra in Fig.~\ref{fig:GeSn_raman_ordered_vs_disordered}(a) is then solely an emergent consequence of the impact of the alloy microstructure on the Raman-active phonon modes in our atomistic calculations. This asymmetric broadening of the Ge-Ge mode peak in Ge$_{1-x}$Sn$_{x}$ was quantified experimentally by Li et al., \cite{Li_APL_2004} who extracted the half-width at half-maximum (HWHM) of the low- and high-frequency sides of their measured Ge-Ge mode Raman peaks for a series of Ge$_{1-x}$Sn$_{x}$ samples. These measured HWHM values were found to be uniformly larger on the low-frequency side of the Ge-Ge Raman peak, with the difference in HWHM between the low- and high-frequency sides increasing with increasing $x$. This behaviour is confirmed by our Sn composition-dependent disordered alloy supercell calculations, where the ratio of the low- and high-frequency Ge-Ge mode HWHM values extracted from computed Raman spectra is found to vary as $\approx 1 + 3.35 \, x$. This implies that the Raman spectrum of pure Ge is approximately symmetric about its peak at frequency $\omega_{\scalebox{0.7}{\text{TO}}} = \omega_{\scalebox{0.7}{\text{LO}}}$ while at, e.g., $x = 10$\% the low-frequency Ge-Ge mode HWHM is expected, on average, to be 1.335 times larger than the high-frequency HWHM. This prediction is in very good quantitative agreement with the measured composition-dependent HWHM values of Ref.~\onlinecite{Li_APL_2004}. Heuristic interpretation of this asymmetric broadening via the alloy's structural properties is straightforward. Lattice relaxation associated with Sn incorporation drives an increase in overall lattice parameter such that, at low $x$ in a relaxed alloy, a majority of the Ge-Ge bonds are slightly stretched with respect to their equilibrium values. This can be observed in the nearest-neighbour relaxed bond length distributions of Fig.~\ref{fig:GeSn_raman_ordered_vs_disordered}(b). In the ordered case (upper panel), relaxation in the presence of a single substitutional Sn impurity creates a bimodal distribution of Ge-Ge nearest-neighbour bond lengths: Ge-Ge bonds close to the Sn atom are compressed by the local lattice expansion about the Sn lattice site, while Ge-Ge bonds further from the Sn atom are stretched (tensile strained) by the global expansion of the lattice parameter. These compressed and stretched bond lengths lie respectively to the left and right of the dashed line in Fig.~\ref{fig:GeSn_raman_ordered_vs_disordered}(b), which denotes the equilibrium Ge-Ge bond length $r_{0}$ (cf.~Table~\ref{tab:vff_parameters}). In the disordered case (lower panel), we again observe that a majority of the Ge-Ge bonds are stretched. This presence of a distribution of tensile-strained Ge-Ge bonds results in a redshift of the individual associated bond optical vibrational frequencies $\omega_{\nu}$ -- again readily understood via the corresponding mode Gr\"{u}neisen parameters $\gamma_{\nu}$, since $\omega_{\nu} \propto \Omega^{-\gamma_{\nu}}$ and $\gamma_{\nu} > 0$ -- giving rise to enhanced broadening and a ``tail'' on the low-frequency side of the Ge-Ge mode peak, with each Ge-Ge bond in the system contributing to the Raman scattering at a frequency determined by the degree to which that bond is stretched or compressed. A majority of the Ge-Ge bonds are stretched, so that their individual mode frequencies are reduced, thereby contributing to enhanced broadening on the low-frequency side of the Ge-Ge Raman peak. Our calculations therefore identify the experimentally observed composition-dependent asymmetric broadening of the Ge-Ge mode Raman spectrum as a direct consequence both of the lattice relaxation, and of the reduction in symmetry associated with the presence of short-range alloy disorder in a substitutional Ge$_{1-x}$Sn$_{x}$ alloy.

Thirdly, we note that the Ge-Sn mode, which is clearly observable in the ordered supercell Raman spectrum in the upper panel of Fig.~\ref{fig:GeSn_raman_ordered_vs_disordered}(a), is significantly broadened and weakened in intensity in the presence of alloy disorder. This finding is consistent with the failure in several experimental studies to resolve a \textit{clear} Ge-Sn feature in the expected frequency range for Sn compositions $x \lesssim 10$\%. \cite{Costa_SSC_2007,Xu_APL_2019,Liu_JRS_2020} Again, interpretation is straightforward in terms of the range of relaxed bond lengths occurring in a disordered Ge$_{1-x}$Sn$_{x}$ alloy. In the ordered supercell the four (compressed) Ge-Sn bond lengths are equal, \cite{Tanner_VFF_2021} and therefore all contribute to Raman scattering at a single frequency. These combined contributions give rise to an identifiable Ge-Sn peak in the alloy Raman spectrum, with the intensity of this peak relative to that of the Ge-Ge peak being roughly -- but not exactly, due to the spread of relaxed Ge-Ge bond lengths -- in proportion to the number of Ge-Sn vs.~Ge-Ge nearest-neighbour bonds in the supercell. In the disordered case, the reduction in symmetry creates a distribution of relaxed Ge-Sn bond lengths, the unequal values of which give rise to contributions to the Raman spectrum across a broad frequency range rather than at a single frequency. Further calculations (not shown) suggest that a clearly visible Ge-Sn feature does emerge in disordered alloy supercell calculations with increasing $x$, as more Ge-Sn bonds form and contribute to the Raman activity, but that this does not occur appreciably for $x \lesssim 10$\%.

Overall, we note that the features associated with optical modes of strained Ge-Ge and Ge-Sn bonds constitute the origin of the ``disorder-activated'' mode described in previous experimental studies, which appears as a long, low-frequency tail on the Ge-Ge mode Raman peak in Ge$_{1-x}$Sn$_{x}$ alloys. \cite{Costa_SSC_2007,Perova_JRS_2017,Bouthillier_SST_2020,Wang_IEEETED_2020}. Qualitatively, our calculations explicitly confirm the origin of this mode as being disorder-activated, and specifically attribute the observed low-frequency tail associated with the Ge-Ge mode to the stretching of Ge-Ge bonds due to Sn-induced lattice relaxation. Quantitatively, our calculations elucidate the nature and evolution of Ge$_{1-x}$Sn$_{x}$ Raman spectrum in terms of individual frequency-dependent Ge-Ge and Ge-Sn bond optical mode contributions to Raman scattering. The presence of alloy disorder gives rise to a distribution of strained nearest-neighbour bond lengths in the alloy, with bond tension (compression) acting to decrease (increase) the associated bond optical vibrational frequencies, giving rise at low $x$ to (i) asymmetric and enhanced low-frequency broadening of the Ge-Ge mode Raman peak, and (ii) broadening and quenching of the Ge-Sn mode peak. As we will describe in Sec.~\ref{sec:results_microstructure} below, this analysis can in certain circumstances allow to obtain direct insight into the alloy microstructure by comparing measured and calculated Raman spectra.


\subsection{Strain-dependent Raman shift}
\label{sec:results_strain}


Having focused above on relaxed Ge$_{1-x}$Sn$_{x}$ alloys, and before turning our attention to further analysis of the features present in our measured and calculated Raman spectra, we note that our theoretical model allows also to predictively parametrise the strain-induced contribution $\Delta \omega ( \epsilon )$ to the Sn-induced shift in the Ge-Ge mode frequency (cf.~Eq.~\eqref{eq:mode_frequency_shift}). Considering the zone-centre TO phonon mode, the strain-induced frequency shift in a pseudomorphic strained layer is given by \cite{Cerdeira_PRB_1972,Rojas-Lopez_JAP_1998,Costa_SSC_2007}

\begin{equation}
    \Delta \omega ( \epsilon ) = \omega_{\scalebox{0.7}{\text{TO}}} \left[ - \gamma_{\scalebox{0.7}{\text{TO}}} \, \text{Tr} ( \epsilon ) + \frac{ 2a_{s} }{3} \left( \frac{ 3 \epsilon_{\perp} - \text{Tr} ( \epsilon ) }{ 2 } \right) \right] \, ,
    \label{eq:strain_dependent_shift_1}
\end{equation}

\noindent
where $\text{Tr} ( \epsilon ) = 2 \epsilon_{\parallel} + \epsilon_{\perp}$ is the trace of the strain tensor, with $\epsilon_{\parallel}$ and $\epsilon_{\perp}$ being the respective components of the strain tensor parallel to the growth plane and along the growth direction; $\omega_{\scalebox{0.7}{\text{TO}}}$ and $\gamma_{\scalebox{0.7}{\text{TO}}}$ are respectively the unstrained TO mode frequency and zone-centre TO mode Gr\"{u}neisen parameter, and $a_{s}$ is the deformation potential describing the impact of an axial strain on the TO mode frequency. The first and second terms in Eq.~\eqref{eq:strain_dependent_shift_1} then respectively describe the impact of the hydrostatic and biaxial components of a pseudomorphic strain on the TO mode frequency. For an [001]-oriented pseudomorphic strained layer $\epsilon_{\perp} = - \frac{ 2 C_{12} }{ C_{11} } \epsilon_{\parallel}$ -- where $\epsilon_{\parallel}$ is the in-plane mismatch between the relaxed lattice parameter of an [001]-oriented Ge$_{1-x}$Sn$_{x}$ pseudomorphic layer and the lattice parameter of the substrate upon which the layer is grown -- such that Eq.~\eqref{eq:strain_dependent_shift_1} yields $\Delta \omega ( \epsilon ) = b \, \epsilon_{\parallel}$ with

\begin{equation}
    b = -2 \omega_{\scalebox{0.7}{\text{TO}}} \left[ \gamma_{\scalebox{0.7}{\text{TO}}} \left( 1 - \frac{ C_{12} }{ C_{11} } \right) + \frac{ a_{s,\scalebox{0.6}{\text{TO}}} }{ 3 } \left( 1 + \frac{ 2 C_{12} }{ C_{11} } \right) \right] \, .
    \label{eq:strain_dependent_shift_2}
\end{equation}

Our DFT calculations then provide all parameters required to compute $b$ analytically (cf.~Eqs.~\eqref{eq:optical_phonon_frequency} --~\eqref{eq:phonon_deformation_potential_axial} and Table~\ref{tab:elastic_parameters}).

As described above, parametrisations of Eq.~\eqref{eq:mode_frequency_shift} based on ``strain-corrected'' experimental data can give rise to significant uncertainty in the composition and strain shift coefficients $a$ and $b$. \cite{Li_APL_2004} This is evidenced by the large spread of $b$ values for Ge$_{1-x}$Sn$_{x}$ reported in the literature (again, summarised in Table I of Ref.~\onlinecite{Gassenq_APL_2017}), which range from as low as $b = 64$ cm$^{-1}$ to as high as $563$ cm$^{-1}$, \cite{Rojas-Lopez_JAP_1998,Lin_APL_2011} with the majority of values estimated based on experimental data being at the higher end of this range. \cite{Fournier-Lupien_APL_2013,Cheng_ECSJSSST_2013,Chang_TSF_2015,Takeuchi_JJAP_2016,Gassenq_APL_2017,Bouthillier_SST_2020} To address the uncertainty associated with this important parameter, we compute $b$ via both analytical and atomistic approaches.

Analytically, we compute the variation of $b$ with $x$ using Eq.~\eqref{eq:strain_dependent_shift_2} in conjunction with the parameters of Table~\ref{tab:elastic_parameters}, yielding $b = 502.9$ cm$^{-1}$ for the Ge-Ge mode in pure Ge. For Sn-containing alloys, we linearly interpolate all parameters appearing in Eq.~\eqref{eq:strain_dependent_shift_2} between those listed in Table~\ref{tab:elastic_parameters} for Ge and $\alpha$-Sn. The resulting calculated $b (x)$ is very close to linear in $x$ for $x \leq 15$\%, fitting to which yields $b (x) = 502.9 + 72.4 \, x$ cm$^{-1}$ and, consequently, $\Delta \omega ( \epsilon ) = 74.2 \, x$ cm$^{-1}$ as the Sn composition-dependent strain dependence of the Ge-Ge mode Raman shift in [001]-oriented pseudomorphic Ge$_{1-x}$Sn$_{x}$/Ge. In the composition range $x \leq 15$\% -- within which most existing Raman spectroscopic data for Ge$_{1-x}$Sn$_{x}$ alloys lies -- our DFT-calculated elastic parameters, combined with Eq.~\eqref{eq:strain_dependent_shift_2}, therefore indicate that $b$ lies in the range $503$ to $518$ cm$^{-1}$.

Atomistically, we compute $b$ using the same set of disordered Ge$_{1-x}$Sn$_{x}$ alloy supercells employed in Sec.~\ref{sec:results_composition}. Firstly, we re-relax these supercells under applied pseudomorphic strain corresponding to growth on an [001]-oriented Ge substrate. This is achieved by constraining the supercell lattice parameter in the (001) plane to remain equal to that of Ge, while allowing the supercell lattice parameter along [001], as well as the internal (ionic) coordinates, to relax. Secondly, we follow the procedure described above to compute the Raman spectra for these pseudomorphically strained Ge$_{1-x}$Sn$_{x}$/Ge supercells, from which we extract the Ge-Ge mode frequency shift $\Delta \omega ( x, \epsilon )$. Finally, for each supercell, we use the computed mode shift $\Delta \omega ( x )$ from the freely relaxed supercell in conjunction with Eq.~\eqref{eq:mode_frequency_shift} to isolate the strain-induced contribution as $\Delta \omega ( \epsilon ) = \Delta \omega ( x, \epsilon ) - \Delta \omega (x)$. Performing a linear fit $\Delta \omega ( \epsilon ) = b \, \epsilon_{\parallel}$ to the resultant data yields $b = 441.8$ cm$^{-1}$ for $x \leq 15$\%.


\begin{figure*}[t!]
	\includegraphics[width=1.00\textwidth]{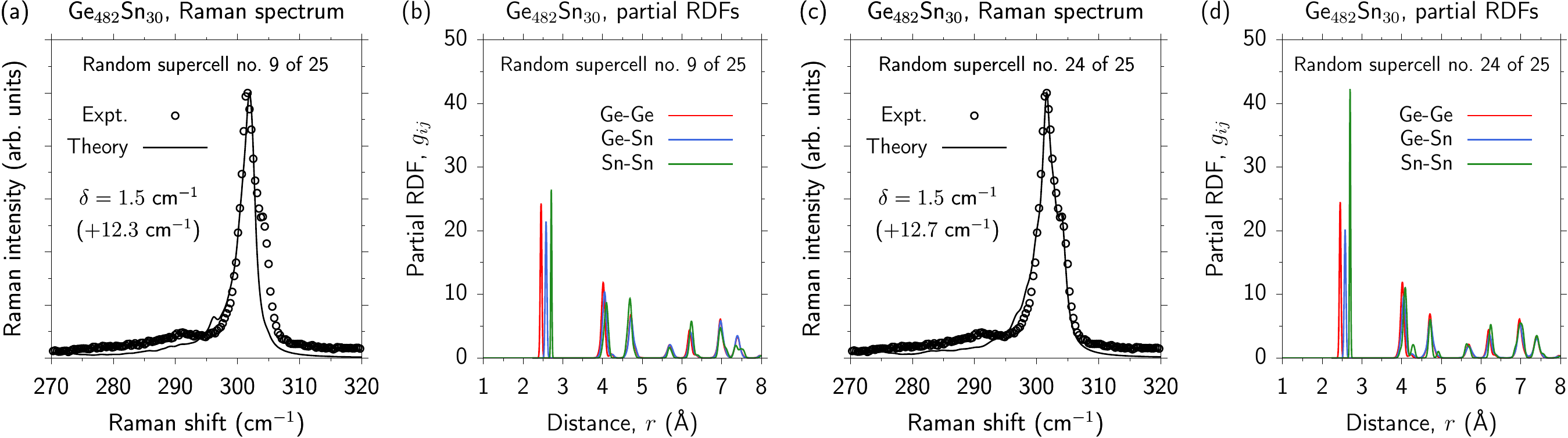}
	\caption{Comparison of calculated Raman spectra and partial RDFs for two of the 25 distinct, disordered Ge$_{482}$Sn$_{30}$ ($x = 5.86$\%) alloy supercells considered in our analysis. (a) and (c) show the calculated Raman spectrum for each disordered supercell (solid black lines), compared to the measured spectrum for the $x = 6$\% NW (open black circles). (b) and (d) respectively show, for the supercells of (a) and (c), the Ge-Ge (solid red line), Ge-Sn (solid blue line), and Sn-Sn (solid green line) partial RDFs. We note that the emergence of a high-frequency ``shoulder'' in the calculated Raman spectrum in (c) correlates with Sn clustering, as revealed by the increased nearest-neighbour Sn-Sn partial RDF in (d).}
	\label{fig:GeSn_raman_spectra_and_partial_rdfs}
\end{figure*}

Our analytical and atomistic calculations of the strain shift coefficient $b$ correspond well to the recent experimental determinations $b = 521$ cm$^{-1}$ by Gassenq et al., \cite{Gassenq_APL_2017} and $b = 491 \pm 52$ cm$^{-1}$ by Bouthillier et al. \cite{Bouthillier_SST_2020} As in the case of the composition shift coefficient $a$ describing the Ge-Ge mode Raman shift, we note that the presence of alloy disorder also acts to decrease the magnitude of $b$ compared to that in an ordered alloy.


\subsection{Correlation between alloy microstructure and Raman spectrum}
\label{sec:results_microstructure}

Finally, we return to the comparison between our calculated and measured Raman spectra in Figs.~\ref{fig:GeSn_theory_vs_experiment}(a) --~\ref{fig:GeSn_theory_vs_experiment}(c). Our calculated spectra in Figs.~\ref{fig:GeSn_theory_vs_experiment}(a) and~\ref{fig:GeSn_theory_vs_experiment}(c) describe well the measured spectra at $x = 1$ and 8\%, capturing quantitatively the asymmetric broadening and emergence of a low-frequency tail (``disorder-activated'' mode) for the Ge-Ge mode at $x = 8$\%. However, the comparison between theory and experiment is less favourable at $x = 6$\%, in Fig.~\ref{fig:GeSn_theory_vs_experiment}(b), where the calculated Raman spectrum fails to capture the measured presence of a shoulder on the high-frequency side of the Ge-Ge mode. Here, we demonstrate theoretically that such a feature can emerge due to the presence of excessive Sn clustering -- relative to the amount of Sn clustering expected in a randomly disordered alloy -- and, as such, can represent an experimental signature of a deviation of the alloy microstructure from the case of a statistically random substitutional alloy.

To this end, it is pertinent to describe what is meant by excessive Sn clustering. In a statistically randomly disordered Ge$_{1-x}$Sn$_{x}$ alloy -- i.e.~in an alloy in which Sn atoms are substituted at Ge lattice sites without bias -- the probability at low $x$ of the occurrence of a Sn-Sn pair formed by substitution of two nearest-neighbour Ge atoms by Sn is $2x^{2}$. In a disordered 512-atom supercell having, e.g., $x = 6.25$\% we therefore expect a total number $512 \times 2 \times (0.0625)^{2} = 4$ of Sn-Sn pairs to occur on average in a randomly disordered alloy. Indeed, for the set of disordered Ge$_{482}$Sn$_{30}$ ($x = 5.86$\%) supercells employed for our comparison to experiment at $x = 6$\%, we find this to be precisely the case: averaging the number of Sn-Sn pairs appearing in each of the 25 distinct, disordered supercells considered gives precisely 4. We note that this is also the case for the distinct sets of 25 alloy supercells employed at all Sn compositions in our analysis, with the average number of Sn-Sn pairs appearing at each Sn composition being $\approx 512 \times 2x^{2}$. This confirms that our analysis incorporates a sufficient number of sufficiently large supercells to accurately describe the structural and vibrational properties of randomly disordered Ge$_{1-x}$Sn$_{x}$ alloys via configurational averaging. Deviation from this probability of occurrence of Sn-Sn nearest-neighbour pairs, as well as from the probabilities of occurrence for larger clusters and complexes of substitutional Sn atoms, then represents a deviation from statistically random alloy disorder. As a consequence, we therefore expect that the calculated Raman spectra of Figs.~\ref{fig:GeSn_theory_vs_experiment}(a) --~\ref{fig:GeSn_theory_vs_experiment}(c) accurately reflect the expected Raman spectra in the case of statistically random alloy disorder, and that any qualitative deviation from these spectral features can be interpreted as corresponding to a deviation from statistically random disorder in the alloy sample on which the measurement was performed.

While the set of supercells employed in our analysis at fixed $x$ contain the expected number of Sn-Sn pairs and larger clusters of substitutional Sn atoms \textit{on average}, this is not necessarily the case for each individual supercell. To examine the appearance in the calculated Raman spectra of features representing deviations from the random alloy (configuration-averaged) spectrum of Fig.~\ref{fig:GeSn_theory_vs_experiment}(b), we therefore consider individually the Raman spectra computed for the 25 distinct, disordered Ge$_{482}$Sn$_{30}$ ($x = 5.86$\%) supercells. For each supercell we associate the calculated Raman spectra to the structural properties by computing the partial radial distribution functions (RDFs). For the atomic species $i$ (= Ge or Sn), the partial RDF $g_{ij} (r)$ describes the net probability of finding an atom of species $j$ (= Ge or Sn) at a distance $r$ from atom $i$. Our calculation of $g_{ij} (r)$ follows Ref.~\onlinecite{Cope_JAC_2007}, where the (spherically averaged) partial RDF is weighted based on the composition of the atomic species $i$ in the supercell. As such, since we are dealing solely with relaxed crystalline structures that maintain a fixed coordination number of four for all atoms, at nearest-neighbour distances the magnitude of the Ge-Ge, Ge-Sn and Sn-Sn partial RDFs is expected to be equal in the case that these nearest-neighbour bonds appear with the number of occurrences expected for a statistically random binary alloy -- i.e.~for a Ge$_{1-x}$Sn$_{x}$ alloy in which the number of each type of nearest-neighbour bond is unambiguously determined by the Sn composition $x$. The results of these calculations are summarised in Fig.~\ref{fig:GeSn_raman_spectra_and_partial_rdfs}, which shows the calculated Raman spectra (solid black lines) and Ge-Ge (solid red line), Ge-Sn (solid blue lines) and Sn-Sn (solid green lines) partial RDFs for two of the 25 disordered Ge$_{482}$Sn$_{30}$ supercells. For comparative purposes, the experimental data of Fig.~\ref{fig:GeSn_theory_vs_experiment}(b) are repeated in Figs.~\ref{fig:GeSn_theory_vs_experiment}(a) and~\ref{fig:GeSn_theory_vs_experiment}(c).

Considering the first of these two supercells, the Raman spectrum and partial RDFs for which are respectively shown in Figs.~\ref{fig:GeSn_raman_spectra_and_partial_rdfs}(a) and~\ref{fig:GeSn_raman_spectra_and_partial_rdfs}(b), we note two key features in the calculated Raman spectrum. Firstly, we note the absence of a shoulder on the high-frequency side of the Ge-Ge mode peak. Secondly, we note the presence of a clear disorder-activated Ge-Ge mode at a frequency of $\approx 296$ cm$^{-1}$, on the low-frequency side of the primary Ge-Ge mode peak. Examining the partial RDFs for this supercell, we note that at nearest-neighbour distances -- i.e.~for $r \lesssim 3$ \AA~-- the magnitudes of the Ge-Ge, Ge-Sn and Sn-Sn partial RDFs are approximately equal. This indicates that the microstructure of this first supercell conforms closely to that expected for a randomly disordered alloy. Indeed, direct inspection of the supercell microstructure reveals the presence of three Sn-Sn nearest-neighbour pairs, with all other Sn atoms having Ge nearest neighbours, and the presence of a single larger Sn cluster consisting of two second-nearest neighbour Sn atoms sharing a common Ge nearest neighbour. The interpretation of the Raman spectrum of Fig.~\ref{fig:GeSn_raman_spectra_and_partial_rdfs}(a) then follows straightforwardly from our discussion in Sec.~\ref{sec:results_disorder}: the emergence of an approximately bimodal distribution of compressed and stretched Ge-Ge bonds in the relaxed supercell respectively give rise to the primary Ge-Ge peak, and to the disorder-activated mode on the low-frequency side of the primary Ge-Ge peak. As such, in this approximately randomly disordered supercell, no feature emerges on the high-frequency side of the Ge-Ge peak.

Turning our attention to the second supercell, the Raman spectrum and partial RDFs for which are respectively shown in Figs.~\ref{fig:GeSn_raman_spectra_and_partial_rdfs}(c) and~~\ref{fig:GeSn_raman_spectra_and_partial_rdfs}(d), we note two key differences. Firstly, in Fig.~\ref{fig:GeSn_raman_spectra_and_partial_rdfs}(c) we note the emergence of a shoulder on the high-frequency side of the Ge-Ge peak in the calculated Raman spectrum, capturing quantitatively that observed in experiment. Secondly, in Fig.~\ref{fig:GeSn_raman_spectra_and_partial_rdfs}(d) we note that the magnitude of the nearest-neighbour Sn-Sn partial RDF exceeds that of Ge-Ge and Ge-Sn by a factor of approximately two. Indeed, inspection of this supercell indicates significant deviation from the microstructure expected for a randomly disordered Ge$_{1-x}$Sn$_{x}$ alloy due to the formation of several large clusters of neighbouring Sn atoms, and complexes involving multiple Sn atoms having shared Ge nearest neighbours. Specifically, this supercell contains a four-atom Sn cluster in the form of a Sn-Sn-Sn-Sn nearest-neighbour chain, as well as a similar three-atom Sn cluster, two Sn-Ge-Sn-Ge-Sn nearest-neighbour chains, and several smaller complexes containing Sn second-nearest neighbours sharing common Ge nearest neighbours. Analysis of the relaxed nearest-neighbour bond lengths in this supercell suggests that the significant local lattice expansion around these large Sn complexes acts to compress nearby Ge-Ge bonds, with the resulting bond compression acting to increase the frequency at which these Ge-Ge bonds contribute to Raman scattering (cf.~Sec.~\ref{sec:results_disorder}). This subset of Ge-Ge bonds, which are compressed to a greater extent than Ge-Ge bonds neighbouring an isolated Sn impurity, then give rise to a clear high-frequency shoulder in the Ge-Ge mode peak. Our analysis then explicitly identifies that the emergence of a high frequency shoulder in the Ge-Ge mode peak of a Ge$_{1-x}$Sn$_{x}$ Raman spectrum can be a signature of the presence of Sn clustering beyond that expected in a randomly disordered alloy. This is a conclusion which we expect to hold generally for defect-free, substitutional Ge$_{1-x}$Sn$_{x}$, and constitutes an explicit link between the atomic-scale alloy microstructure and the experimentally measurable Raman spectrum.


\section{Conclusions}
\label{sec:conclusions}

In summary, we have presented a combined theoretical and experimental analysis of the structural and vibrational properties of Ge$_{1-x}$Sn$_{x}$ alloys based on Raman spectroscopy, providing for the first time detailed atomistic theoretical insight into the features and evolution of the Raman spectra of this emerging group-IV material system.

Experimentally, we reported VLS (LICVD) growth of Ge$_{1-x}$Sn$_{x}$ NWs having Sn compositions $x \leq 8$\%, for which substitutional Sn incorporation, compositional uniformity and high crystalline quality were confirmed via SEM, STEM and EDX characterisation. Raman spectroscopic measurements were performed on these NW samples at fixed temperature, incident optical excitation power, and incident optical polarisation. The fully relaxed nature of the free-standing NWs allows explicit access to the pure Sn composition dependence of the widely investigated Ge-Ge mode optical vibration frequency.

Theoretically, we presented a fully analytic anharmonic VFF potential tailored to simulate the vibrational properties of Ge$_{1-x}$Sn$_{x}$ alloys. The VFF potential was directly parametrised using a combination of DFT-calculated second- and third-order relaxed and inner elastic constants, allowing to describe exactly the zone-centre optical phonon frequencies and their associated mode Gr\"{u}neisen parameters. The low computational cost associated with this VFF potential provides the ability to perform high-throughput calculations for realistic (large, disordered) alloy supercells. We further note that the VFF potential presented in this work is broadly applicable to investigate the structural, elastic and vibrational properties of cubic-structured, isovalent multinary alloys (e.g.~Ge$_{1-x}$C$_{x}$, Ge$_{1-x}$Pb$_{x}$ or Si$_{y}$Ge$_{1-x-y}$Sn$_{x}$) and, via inclusion of appropriate electrostatic terms, \cite{Tanner_PRB_2019} can be extended to treat cubic-structured polar alloys (e.g.~III-V semiconductor alloys).

Using the VFF potential we performed lattice relaxation and computed phonon modes for Ge$_{1-x}$Sn$_{x}$ alloy supercells, allowing direct atomistic computation of Raman spectra for comparison to, and interpretation of, experimental measurements. Detailed analysis demonstrated the ability of this approach to accurately describe the evolution of the Raman spectrum with Sn composition $x$, and quantified several key features observed in previous experimental studies. Our calculations suggest that the magnitude of the Sn composition-dependent Raman shift of the primary Ge-Ge mode feature varies depending on the presence or absence of short-range ordering, potentially accounting in part for the significant spread in values of the composition shift coefficient $a$ reported in the literature. We further demonstrated that the failure to observe clear features related to Ge-Sn vibrational modes in measured Raman spectra for $x \lesssim 10$\% is a consequence of short-range alloy disorder. We also elucidated the origin of the ``disorder activated'' mode described in previous experimental studies in terms of the distribution of relaxed Ge-Ge nearest-neighbour bond lengths in the alloy. Based on our DFT-calculated bulk elastic properties and atomistic VFF alloy supercell calculations, we computed -- analytically and atomistically -- the strain shift coefficient $b$. Finally, we demonstrated correlation between features in measured Raman spectra and the alloy microstructure, specifically showing that the presence of a high-frequency ``shoulder'' on the primary Ge-Ge Raman peak can be interpreted as a signature of clustering of substitutional Sn atoms beyond that expected in a randomly disordered Ge$_{1-x}$Sn$_{x}$ alloy.

Overall, our combined theoretical and experimental analysis provides significant new insight to aid the characterisation of Ge$_{1-x}$Sn$_{x}$ alloys via Raman spectroscopy, narrowing the range of parametric uncertainty associated with determining the alloy composition and strain from measured Raman spectra, while also providing a calculational framework that potentially allows to correlate signatures in measured Raman spectra directly to the microstructure of a given alloy sample, thereby providing useful information for characterisation and further analysis of this emerging material system.


\section*{Acknowledgements}

This work was supported by Science Foundation Ireland (SFI; project nos.~15/IA/3082 and 14/IA/2513), by the Science and Engineering Research Board, India (SERB; project no.~EMR/2017/002107), by the Irish Research Council (IRC; via Government of Ireland Postgraduate Scholarship no.~GOIPG/2015/2772, held by J.D.), and by the National University of Ireland (NUI; via the Post-Doctoral Fellowship in the Sciences, held by C.A.B.).



%

\end{document}